\documentclass[10pt,journal,twocolumn,twoside]{IEEEtran} 
\IEEEoverridecommandlockouts
\usepackage{setspace}
\usepackage{subfigure}
\usepackage{graphicx}
\usepackage{epstopdf}
\usepackage{float}
\usepackage{amsmath}
\usepackage[ruled]{algorithm2e}
\usepackage{algpseudocode}
\usepackage{array}
\usepackage{amsthm}
\usepackage{amsmath}
\usepackage{amssymb}
\usepackage{mdwmath}
\usepackage{mdwtab}
\usepackage{eqparbox}
\usepackage{stfloats}
\usepackage{fixltx2e}
\usepackage{cases} 
\usepackage{caption}
\usepackage{graphicx}
\usepackage{float} 
\usepackage{flushend}

\usepackage{xcolor}
\usepackage{makecell}

\usepackage{url}
\usepackage{cite}
\allowdisplaybreaks[4]

\newtheorem{lemma}{\bf Lemma}

\newtheorem{remark}{Remark}

\usepackage{wrapfig}

\newcommand{\lemmaref}[1]{Lemma~\ref{#1}}

\newcommand{\appref}[1]{Appendix~\ref{#1}}

\newcounter{MYtempeqncnt} 
\makeatletter
\renewcommand{\maketag@@@}[1]{\hbox{\m@th\normalsize\normalfont#1}}%
\makeatother

\begin{document}

\title
{Near-Field Integrated Sensing and Communication: Performance Analysis and Beamforming Design}

\author{ Kaiqian Qu,~\IEEEmembership{Student Member, IEEE}, Shuaishuai Guo,~\IEEEmembership{Senior Member, IEEE}, \\and Nasir Saeed,~\IEEEmembership{Senior Member, IEEE}
\thanks{The work is supported by in part by the National Natural Science Foundation of China under Grant 62171262; in part by Shandong Provincial Natural Science Foundation under Grant ZR2021YQ47; in part by the Taishan Young Scholar under Grant tsqn201909043; in part by Major Scientific and Technological Innovation Project of Shandong Province under Grant 2020CXGC010109; and in part by Shandong Nature Science Foundation under Grant ZR2021LZH003. (\emph{Corresponding Author: Shuaishuai Guo})}
\thanks{ 
K. Qu and S. Guo are  with School of Control Science and Engineering, Shandong University, Jinan 250061, China, and also with Shandong Key Laboratory of Wireless Communication Technologies, Shandong University, China (e-mail: qukaiqian@mail.sdu.edu.cn,shuaishuai\_guo@sdu.edu.cn).
N. Saeed is with the Department of Electrical and Communication Engineering, United Arab Emirates University (UAEU), Al Ain, United Arab Emirates (e-mail: mr.nasir.saeed@ieee.org).}
   }
\maketitle


\begin{abstract}
This paper explores the potential of near-field beamforming (NFBF) in integrated sensing and communication (ISAC) systems with extremely large-scale arrays (XL-arrays). The large-scale antenna arrays increase the possibility of having communication users and targets of interest in the near field of the base station (BS).  The paper first establishes the models of electromagnetic (EM) near-field spherical waves and far-field plane waves. With the models, we analyze the near-field beam focusing ability and the far-field beam steering ability by finding the gain-loss mathematical expression caused by the far-field steering vector mismatch in the near-field case. We formulate the NFBF design problem as minimizing the weighted summation of radar and the communication beamforming errors under a total power constraint and solve this quadratically constrained quadratic programming (QCQP) problem using the least squares (LS) method. Moreover, the Cramér-Rao bound (CRB) for target parameter estimation is derived to verify the performance of NFBF. Furthermore, we also perform power minimization using convex optimization while ensuring the required communication and sensing quality-of-service (QoS). The simulation results show the influence of model mismatch on near-field ISAC and the performance gain of transmit beamforming from the additional distance dimension of near-field.
\end{abstract}


\section{Introduction} 
The development of mobile communication technology has led to explosive growth in communication devices and services. It is estimated that by 2025, the number of communication devices will increase to 25 billion. However, limited spectrum resources are making it difficult to cope with the growing demand for communications. To meet this demand, communication devices are being forced to share frequency bands with other devices\cite{Liu2019JointRA}. To address this issue, the increase in communication bands and the development of multi-input multi-output (MIMO) technology have made the similarity between communication and radar systems more obvious regarding spectrum applications, hardware devices, and transmission technology. As a result, academia and industry consider the convergence of communication and sensing (radar)\footnote{In this paper, we consider that radar and sensing have the same meaning.} as an important technology evolution direction for the sixth generation (6G) communications  \cite{Liu2019JointRA,Liu2021IntegratedSA}. Meanwhile, the increasing size of the antenna is a realistic trend. In this context, 
extremely large-scale array (XL-array) and integrated sensing and communication (ISAC) functions are the future trends of the 6G base station (BS)\cite{Saad2019AVO}. 
\subsection{Prior Works on ISAC}
In the earlier studies, communication radar spectrum sharing (CRSS) was the most discussed ISAC model, which enables the coexistence of communication and perception by using opportunity spectrum sharing and embedding communication symbols in sensing waveform \cite{Liu2017MUMIMOCW,Saruthirathanaworakun2012OpportunisticSB,Liu2018MIMORA,Zheng2019RadarAC,9241739}. However, unlike CRSS, ISAC aims to fuse communication and sensing, sharing not only spectrum but also hardware resources. In the ISAC system, the BS uses multi-antenna technology to achieve target sensing while simultaneously downlinking communication through joint transmit beamforming. In fact, shared deployment performs better for ISAC than for CRSS  \cite{Liu2017MUMIMOCW}.  Therefore, this fully shared ISAC system is now widely adopted by researchers.

At present, much research has been done in the field of transmit beamforming or waveform design of multi-antenna ISAC system\cite{Liu2017TowardDR,Liu2018HybridBW,Tsinos2021JointTW,Luo2020MultibeamOF,Liu2021CramrRaoBO,Liu2019JointTB,Cong2023VehicularBB,9850347}. In \cite{Liu2017TowardDR}, the authors studied transmit
beamforming in ISAC systems with a multi-user MIMO (MU-MIMO) setup, in which the communication waveform is utilized as a radar transmit waveform. A beampattern matching method was proposed to design the sensing function while reducing the multi-user interference (MUI) to achieve the purpose of integrated waveform design. Furthermore, \cite{Liu2017TowardDR} proposed a tradeoff design method that adds a compromising factor to make the design more flexible. Similar ideas have been adopted in many recent works \cite{Liu2018HybridBW,Tsinos2021JointTW}. For example, in \cite{Liu2018HybridBW}, a hybrid beamforming design for a millimeter wave (mmWave) band ISAC system was considered using the tradeoff ideas by formulating the objective of minimizing the weighted summation of the communication and radar beamforming errors and solving the non-convex problem.
The authors in \cite{Liu2021CramrRaoBO} mainly considered optimizing waveform parameter estimation performance under the condition of meeting communication needs, in which the user's signal-to-interference-plus-noise ratio (SINR) threshold and Cramér-Rao bound (CRB) are respectively used as communication and sensing performance measures. In \cite{Liu2019JointTB}, the authors designed transmit beamforming to optimize both the radar transmit beam pattern
and the SINR at the communication users. In this case, the transmitter utilizes individual communication and radar waveform for joint beamforming, thus increasing the degree of freedom (DoF) of the MIMO radar. Applying ISAC to vehicular networks,  \cite{Cong2023VehicularBB} proposed a vehicular
behavior-aware ISAC beamforming design.
 
The above studies generally assumed that both the communication user and the target are in the far field of the antenna array, i.e., the distance between them is greater than the Fraunhofer distance \cite{Selvan2017FraunhoferAF}. 
In traditional communication or radar sensing, the scale of the antenna array is relatively small due to the limitation of the antenna size. Therefore, the near-field region is small, and the assumption of far-field plane waves is applicable. However, with the development of ISAC systems, such as mmWave communications and radar, terahertz (THz) communications and imaging, BS may be equipped with XL-array \cite{Cui2022NearFieldCF} or large intelligent surfaces (LISs) \cite{Torres2020NearAF}. This will increase the array aperture and a decrease in wavelength, causing the near-field region of BS to grow rapidly, leading to an increased possibility of sensing and communication occurring in the near field. Since the EM wavefront exhibits a spherical shape in the near field,  the related study should be conducted under the spherical wave model.

Research on the near field started earlier in the field of radar signal processing, with a focus on localizing signal sources \cite{Huang1991NearfieldMS,Swindlehurst1988PassiveDA,Tao2011JointDR}. Recently, there has been a growing interest in studying communication in the near field of large-scale antennas \cite{Zhou2015SphericalWC,Cui2021ChannelEF,Wu2021DistanceAwarePF,Lu2022NearFieldMA,Lu2021CommunicatingWE,Zhang2021BeamFF}. For instance, Zhou et al. proposed an analytical spherical-wave channel model for large linear arrays in \cite{Zhou2015SphericalWC}. Their study demonstrated that spherical waves can help decorrelate channels, improving the ability to distinguish between different users. In \cite{Cui2021ChannelEF}, a polar-domain representation for the near-field channel was proposed that accounts for both the angular and distance information. In \cite{Wu2021DistanceAwarePF}, it was shown that the near field has a DoF, which is proportional to the product of the BS and user equipment (UE) array apertures and inversely proportional to the BS-UE distance. Moreover, Lu et al. performed a unified modeling and analysis of the XL-MIMO system in \cite{Lu2022NearFieldMA,Lu2021CommunicatingWE}, demonstrating the importance of correctly modeling the near field. In \cite{Zhang2021BeamFF}, Zhang et al. investigated near-field communication with different antenna structures, including dynamic metasurface antennas (DMA), and demonstrated the near-field beam focusing capability can mitigate interference in both angular and distance domains, leading to improved transmission rates. The existing studies on the near-field communication suggest that the propagation characteristics of electromagnetic waves in the near field can significantly impact the performance of communication and sensing systems \cite{Zhang2021BeamFF,Huang1991NearfieldMS,Swindlehurst1988PassiveDA,Tao2011JointDR,Zhang2018LocalizationON,Zhou2015SphericalWC,Cui2021ChannelEF,Wu2021DistanceAwarePF,Lu2022NearFieldMA,Lu2021CommunicatingWE}.

However, to the best of our knowledge, the research on such a near-field ISAC system has not been well studied in the literature under the trend of continuous expansion of the near-field region \cite{Qu2023NearFieldIS}. This motivates us to look into accurate modeling, ISAC beamforming, and performance analysis in the near-field region.
\subsection{Contributions}
Specifically, this paper investigates the near-field accurate modeling, the NFBF design, and performance analysis for the ISAC system with XL-array.
The contributions of this paper are summarized as follows:
\begin{itemize}
\item First, we model the near-field  spherical wave and far-field plane wave and analyze their differences. Additionally, we model the near-field ISAC system and point out the advantage of near field for ISAC. 
\item Then, we jointly design NFBF using the method proposed in \cite{Liu2018HybridBW} to approximate both fully digital communication and radar beamformers. We also introduce an auxiliary unitary matrix to resolve the problem of unequal targets and users. Our solution algorithm is faster and performs comparably to the alternating minimization (AM) algorithm in \cite{Liu2018HybridBW}. 
\item After that, we introduce the two-dimension multiple signal classification (2-D MUSIC) algorithm for sensing parameter estimation and derive the CRB for 2-D parameter estimation to analyze the beam performance.
\item Finally, to investigate the power-saving potential of NFBF while ensuring communication and sensing QoS, we focus on the power minimization problem and employ the semidefinite relaxation (SDR) technique to transform it into a convex optimization problem. Our simulation results demonstrate that NFBF has better parameter estimation and higher communication rates when the target or user is in the near field, highlighting the focusing capability of the near-field beam. And with the decrease of target distance, near-field ISAC will have higher estimation performance. We also verify that the near-field ISAC beam design will reduce the power. The simulations further emphasize the importance of proposed accurate modeling for XL-arrays.
\end{itemize}

\subsection{Organization and Notation}
The remainder of this paper is structured as follows. In Section II, we present the system model, which includes near-field and far-field models. In Section III, we address the near-field integrated beam trade-off design problem under the total power constraint. Section IV introduces the method of echo parameter estimation and derives CRB. Section V presents the design of a beam with minimum power that meets both communication and sensing requirements. In Section VI, we provide simulation results and analysis. Finally, in Section VII, we conclude this paper.

\emph{Notations:} We use boldface lowercase and uppercase letters for vectors and matrices, respectively. Subscripts indicate the location of the entry in the matrices or vectors, i.e., $s_{i,j}$ and $x_n$ are the $(i,j)$-th and the $n$-th element in $\mathbf{S}$ and $\mathbf{x}$. 
$\| \cdot \|$,  $\| \cdot \|_F$ denote the $\ell_2$ norm and Frobenius norm.
$(\cdot)^T$, $(\cdot)^H$, $(\cdot)^*$ and $(\cdot)^{\dag}$ stand for transpose, Hermitian transpose, complex conjugate and Moore-Penrose pseudo-inverse of the matrices, respectively. The vectorization operator, trace, Kronecker product, and stochastic expectation are written as 
${\rm vec}(\cdot)$, ${\rm {Tr}}\left(\cdot\right)$, $\otimes$, and $\mathbb{E}\{\cdot\}$. 
The notations $\operatorname{Re}\{\cdot\}$ and $\operatorname{Im}\{\cdot\}$ denote the real and imaginary part of a complex number, respectively. 

\section{System Model}
In this section, we consider ISAC systems where the BS is equipped with XL-array. 
We first introduce the near-field and far-field models in Section II-A to facilitate the subsequent study. Then, we perform a gain analysis of the array response vector in the near and far fields. After that, we describe the detailed configuration of the ISAC system in Section II-C. Finally, we establish the communication transmission model and the sensing model of the ISAC system. 

\subsection{Near-Field and Far-Field Models}
\begin{figure}[htbp]
    \centering
    \includegraphics[width=1\linewidth]{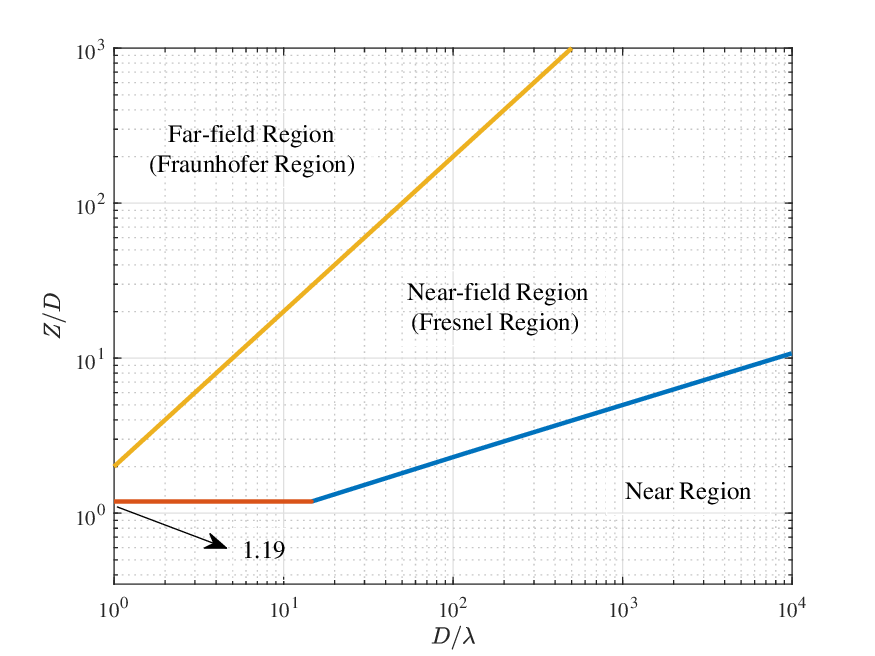}
    \caption{The boundaries of different regions.}   
    \label{fig2a}
\end{figure}

Although there are no strict EM distance boundaries between EM fields, different fields correspond to different approximate structures. According to \cite{1137900}, the diffraction field is usually divided into three regions: Near Region, Fresnel Region (near-field) and Fraunhofer Region (far-field). 
The boundary between the far field and the Fresnel region is usually given by Fraunhofer distance $d_F$ which is obtained when the maximum allowable phase error is $\frac{\pi}{8}$, 
\begin{equation} \label{eq1}
 d_\mathrm{F}=\frac{2\,D^2}{\lambda},
 \end{equation} 
where $D$ is the antenna diameter and $\lambda$ represents the wavelength. The same condition will be used to determine both the maximum phase and amplitude differences for the lower Fresnel region boundary (see [31, Eq. (5-6)]). The boundaries of different regions is shown in Fig. \ref{fig2a}. Note that the near region is very small and can be ignored, so the 'near field' mentioned in this paper refers to the Fresnel Region.

\begin{figure*}[htbp]
\centering
\subfigure[]{{\label{plane}}\includegraphics[width=0.3\linewidth]{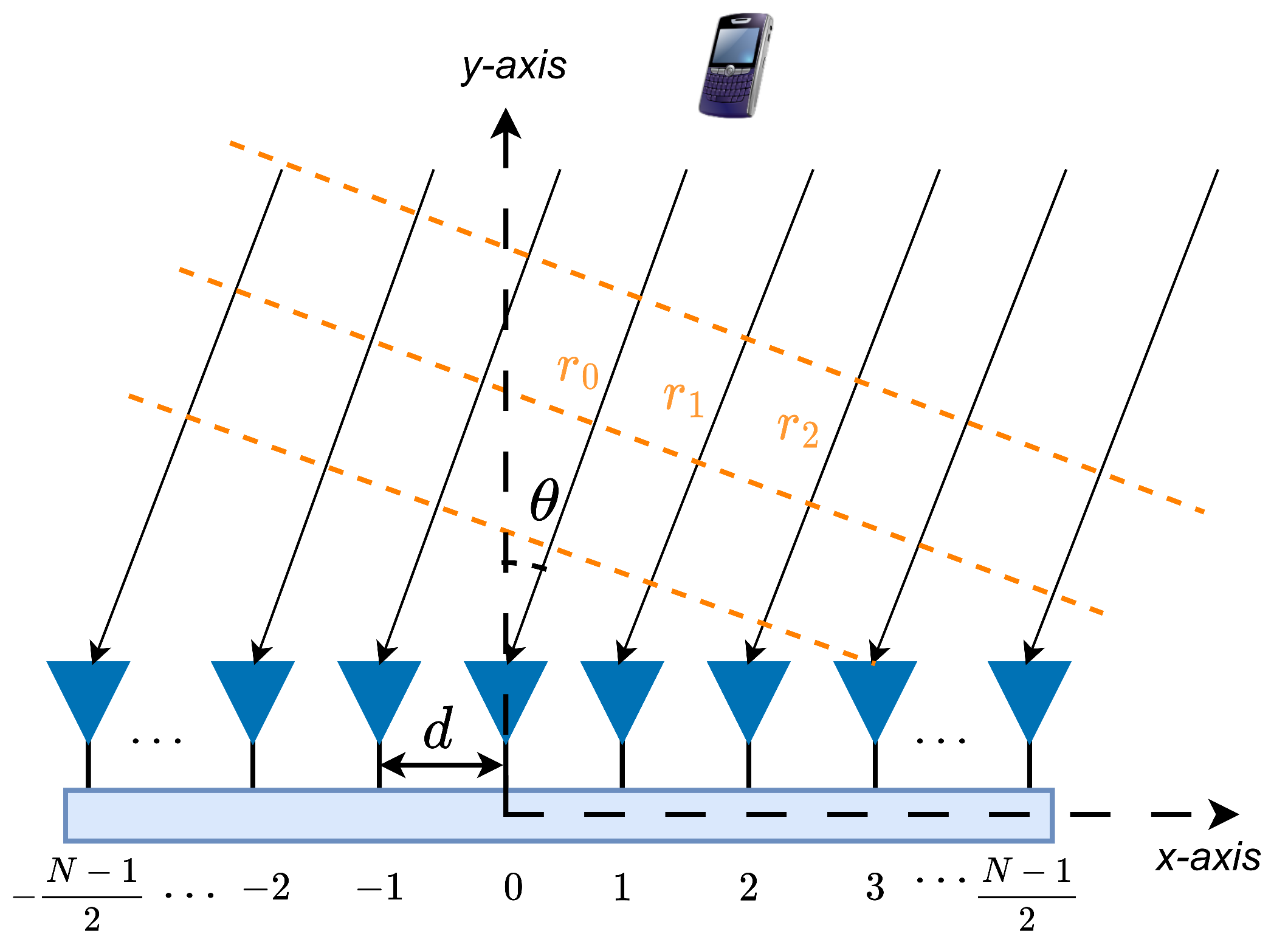}}
\subfigure[]{{\label{speher}}\includegraphics[width=0.3\linewidth]{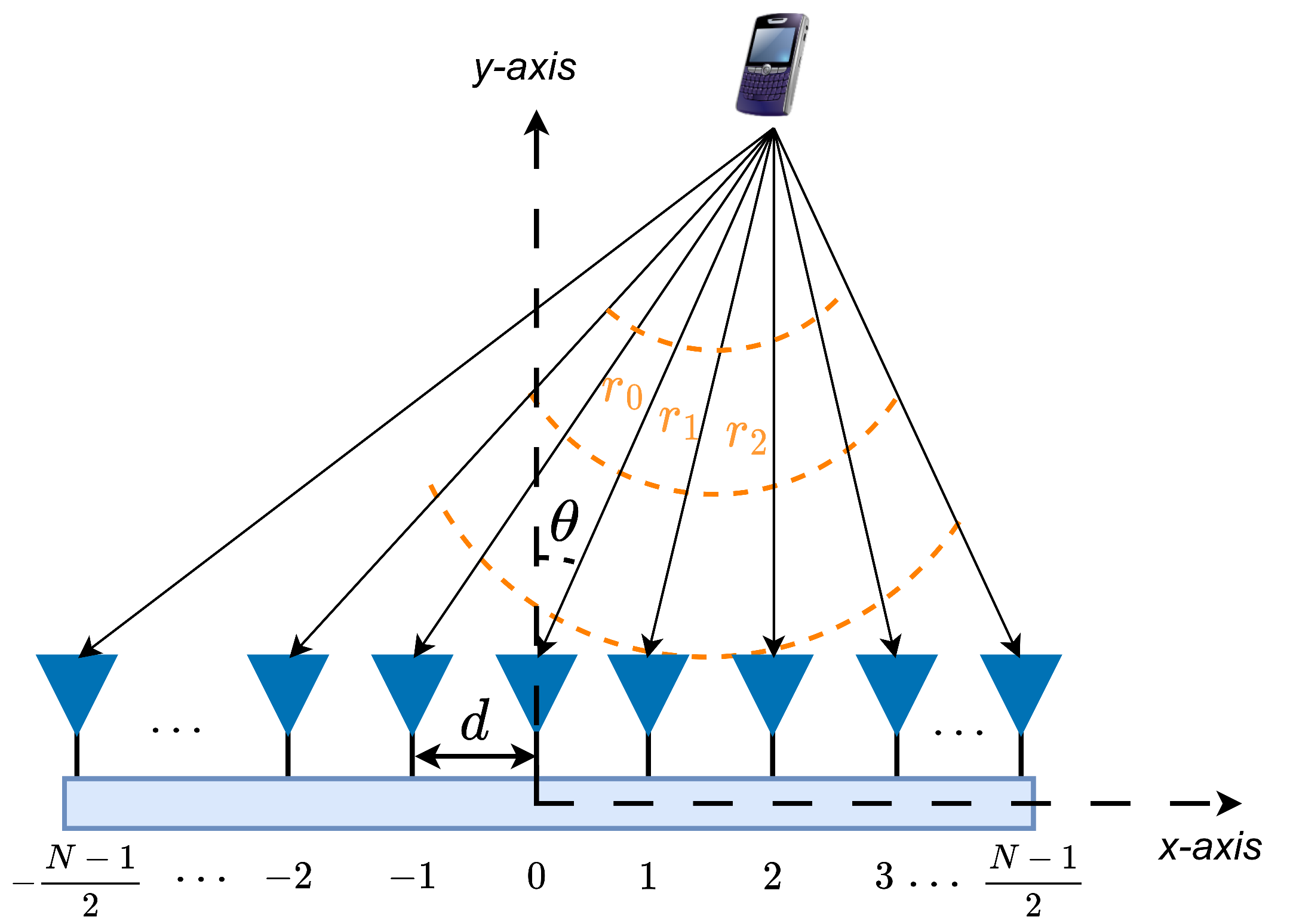}}
\subfigure[]{{\label{fig4}}\includegraphics[width=0.33\linewidth]{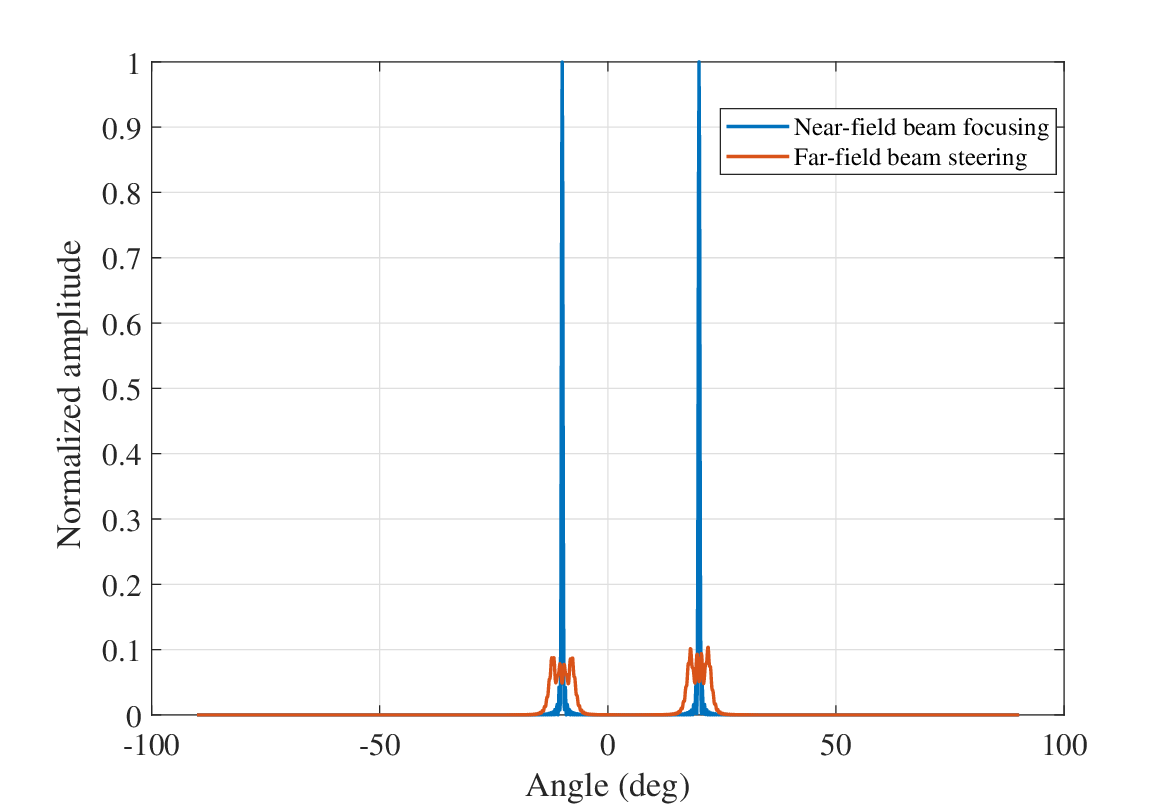}}
\caption{(a)-(b): Plane wave (a) and spherical wave (b) models for ULA; (c): Comparison of near-field beam focusing and far-field beam steering at 10 m from a 256-antenna BS when the carrier is 28 GHz.}
\end{figure*}

When in the far field, the EM wave is transmitted as an approximated plane wave, in the case of the uniform linear array (ULA), as shown in Fig. \ref{plane}. 
The phase difference generated by the signal at the different receiving antennas can be expressed in the form of a steering vector\footnote{This paper distinguishes between the near-field focusing vector $\mathbf{a}\left(r,\theta\right)$ and the far-field steering vector $\mathbf{a}\left(\theta\right)$ by the number of variables in the function vector $\mathbf{a}$.},
\begin{equation}\label{eq3}
\begin{split}
    \mathbf{a}\left(\theta\right)&=\big[e^{j2\pi \frac{(-N+1)d\sin{\theta}}{2\lambda}}, e^{j2\pi \frac{(-N+3)\sin{\theta}}{2\lambda}},\ldots, e^{j2\pi \frac{(N-1)\sin{\theta}}{2\lambda}}\big]^T,
\end{split}
\end{equation}
where $f$ represents the carrier frequency, $d$ denotes the antenna spacing. When the antenna is distributed at half wavelength i.e. $d=\frac{\lambda}{2}$, (\ref{eq3}) can be simplified as $\big[e^{j\pi \frac{(-N+1)}{2} \sin{\theta}}, \ldots, e^{j\pi \frac{(N-1)}{2}\sin{\theta}}\big]^T$.

However, in future 6G XL-MIMO systems, Fraunhofer distance increases sharply with the increase of antenna size and working frequency band, just as shown in Fig. \ref{fig2a}. For example, an antenna array with array aperture $D=2 \rm{~m}$ working at 28 Ghz has a near-field coverage distance of $746.7 \rm{~m}$, which is a large range that can cover many user equipments (UEs) and targets. Thus, the near-field region in XL-MIMO becomes not negligible,  and the original plane wave assumption no longer applies. 
In the near field, the  spherical wave model is shown in Fig. \ref{speher}. 

According to the triangular edge relationship, the distance from the $n$-th antenna to the signal source is $r_{n}=\sqrt{r^2+(\delta_nd)^2-2r\delta_nd\sin{\theta}}$, where $\delta_n=\frac{2n-N+1}{2}$, $r$ and $\theta$ denotes the distance and the angle from the signal source to the reference array element, i.e. the array element numbered $\bf{0}$ in Fig. \ref{speher}. Disregard the signal amplitude difference at different antennas in the Fresnel region \cite{1137900}. Then we denote the near-field focusing vector as
\begin{equation}\label{eq4}
    \mathbf{a}(r,\theta)=\big[e^{-j2\pi \frac{r_{1}-r}{\lambda}},e^{-j2\pi \frac{r_{2}-r}{\lambda}},\ldots, e^{-j2\pi \frac{r_{N}-r}{\lambda}}\big]^T.
\end{equation}

Note that the properties of the near field are encapsulated in (\ref{eq4}). In the near field, as shown in Fig. \ref{fig4} beam steering using (\ref{eq3}) have similar properties to beam focusing using (\ref{eq4}) in the angular domain but with much less gain than the latter, because far-field steering vector (\ref{eq3}) is an approximation of near-field focusing vector (\ref{eq4}) at an infinite distance and a mismatch will occur in the near field. 




\subsection{Gain Loss Analysis of Far-field Steering Vector}
\begin{figure}[!h]
    \centering
    \includegraphics[width=1\linewidth]{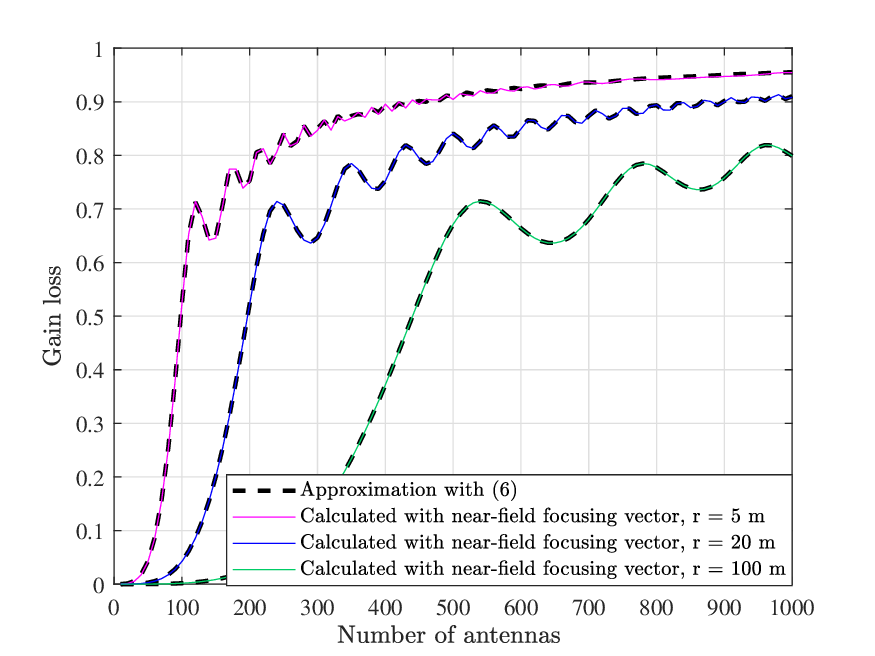}
    \caption{ Gain loss of far-field steering vector at different distances.}
    \label{gain}
    \end{figure}
The far-field steering aptitude approximates the near-field focusing vector at a sufficient distance. Intuitively, due to the mismatch of the far-field steering vector in the near-field region, the original far-field beamforming (FFBF) method will inevitably result in a loss of gain. Then the normalized gain loss can be expressed as
\begin{equation}
    \label{gainloss}
    l=1-\frac{|\mathbf{a}^H(\theta)\mathbf{a}(r,\theta)|}{|\mathbf{a}^H(r,\theta)\mathbf{a}(r,\theta)|}.
\end{equation}
We have the following lemma based on the gain loss expression in (\ref{gainloss}).
\begin{lemma}
\label{lemma1}
\rm{The gain loss caused by the far-field steering vector mismatch can be approximated as follows }
\begin{equation}\label{obb}
l\approx1-\frac{1}{N}\sqrt{2N+8\sum_{i=1}^{\frac{N}{2}-1}\sum_{j=1}^{\frac{N}{2}-1}\cos\left(\frac{\lambda \pi}{4r}\cos^2{(\theta)}\mathbf{W}\right)_{i,j}},
\end{equation}
where $r$ and $\theta$ are the distance and angle to the center point of the antenna array. $\mathbf{W}$ is a $\left(\frac{N}{2}-1\right) \times \left(\frac{N}{2}-1\right)$ constant upper triangular matrix, the element of which ${w}_{i,j},\ {i,j}\in\{1,2,\cdots,\frac{N}{2}-1\}$ can be expressed as 
\begin{equation}\label{w}
        {w}_{i,j}=\left\{
\begin{aligned}
&\left(i-1\right)^2-j^2+\left(N-1\right)\left(j+1-i\right), &i\le j\\
&\quad\quad\quad\quad\quad\quad\quad0\  \quad\quad\quad\quad\quad\quad\quad, &i>j
\end{aligned}.
\right.
\end{equation}
\end{lemma}
\begin{IEEEproof}
    See in the \appref{A1}.
\end{IEEEproof}
\begin{remark}
\rm Without loss of generality, \lemmaref{lemma1} is derived under the assumption that the number of antennas $N$ is even. When $N$ is odd, a similar conclusion can be obtained with only minor changes in the derivation.
\end{remark}
\subsection{ISAC System}
\begin{figure}[!h]
       \centering
       \includegraphics[width=1\linewidth]{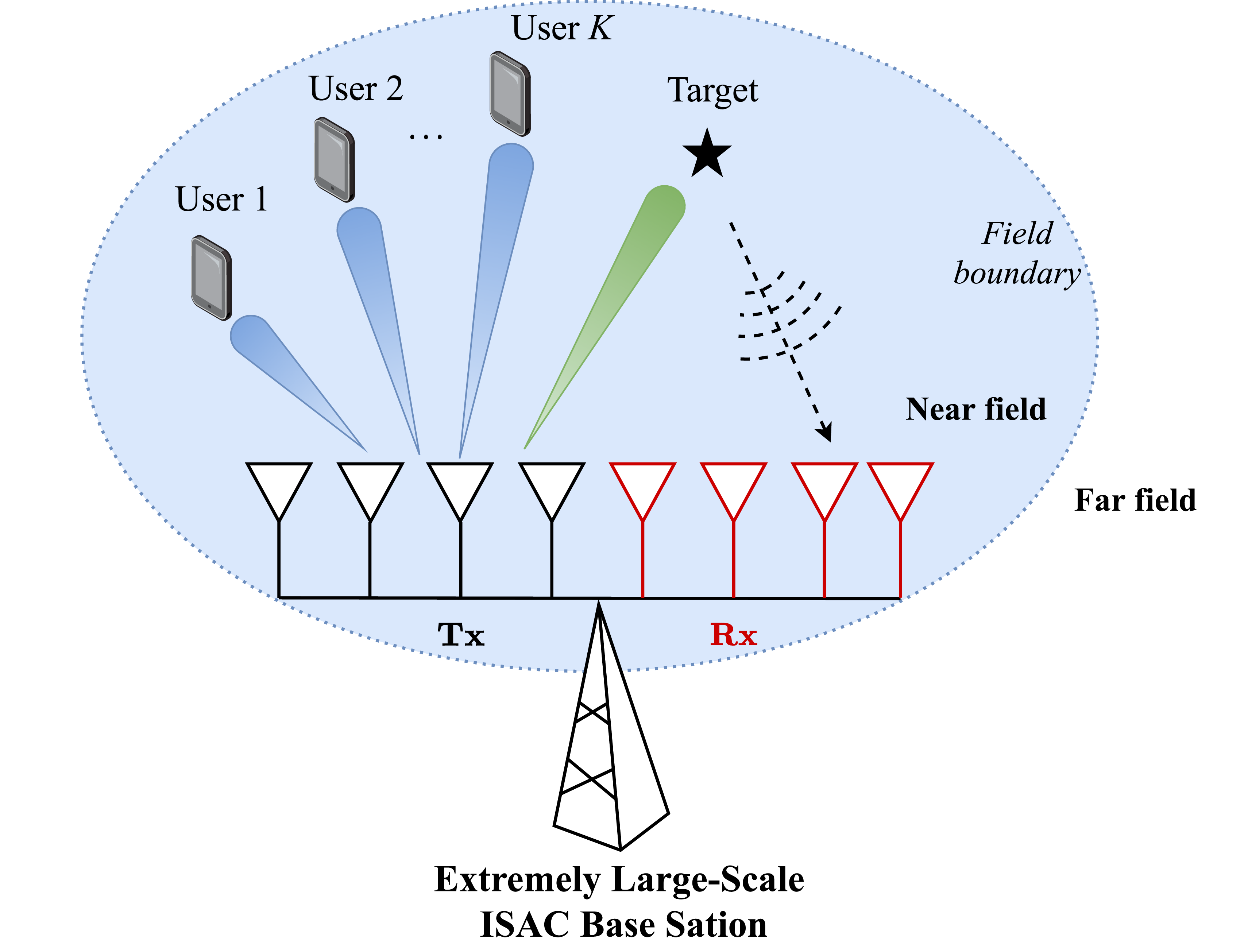}
       \caption{System model of near-field ISAC.}
       \label{System Model}
\end{figure}
This paper considers the ISAC system shown in Fig. \ref{System Model}, where the BS is equipped with XL-array, including $N_t$ transmitting antennas and $N_r$ receiving antennas.  The transmitting antenna transmits waveforms, serving $K$ single-antenna users while detecting a single target at the same time.  The receiving antenna receives echoes generated by the target. Both transmitting and receiving antennas are distributed in a uniform line and spaced at half-wavelength intervals.

\subsubsection{Communication Model}
 For communication services, we consider the following model
\begin{equation}
    \label{eq6}
\mathbf{Y}_C=\mathbf{H}\mathbf{F}\mathbf{S}+\mathbf{N},
\end{equation}
where $\mathbf{Y}_C\in\mathbb{C}^{K \times L}$ represents the matrix of symbols received at the users with $L$ being the length of symbols at discrete time, $\mathbf{H}=\big[\mathbf{h}_1, \mathbf{h}_2, \ldots, \mathbf{h}_K\big]^H\in\mathbb{C}^{K \times N_t}$ is the near-field channel matrix, $\mathbf{F}=\big[\mathbf{f}_1, \mathbf{f}_2, \ldots, \mathbf{f}_{K}\big]\in\mathbb{C}^{N_t \times K}$ stands for the beamforming matrix with $N_s$ being the number of data streams\footnote{Without losing generality, this paper assumes that $N_s=K$.}, $\mathbf{S}\in\mathbb{C}^{N_s \times L}$ is the transmitted symbol matrix, which satisfies that $\mathbb{E}(\mathbf{S}\mathbf{S}^H)=\mathbf{I}_{N_s}$, and finally $\mathbf{N}\in\mathbb{C}^{K \times L}$ is independent, zero-mean complex Gaussian noise matrix with known covariance $\mathbf{R}_N=\sigma_n^2\mathbf{I}_K$.
For the ISAC model, let 
\begin{equation}
    \label{eq7}    \mathbf{X}=\mathbf{F}\mathbf{S}\in\mathbb{C}^{N_t \times L},
\end{equation}
where \textbf{X} is the dual-functional waveform for both sensing and communication. Also, we assume that the near-field channel \textbf{H} remains unchanged during the length of symbols and can be perfectly estimated by near-field channel estimation methods in\cite{Cui2021ChannelEF}. Following \cite{Cui2021ChannelEF}, the spherical wave channel model in the near field is presented as 
\begin{equation}
    \label{eq8}
    \mathbf{h}_k^{near}=\alpha_0\mathbf{a}(r_0,\theta_0)+\sqrt{\frac{1}{P}}\sum_{l=1}^P\alpha_l\mathbf{a}(r_l,\theta_l),
\end{equation}
where $P$ is the number of near-field non-line-of-sight (NLoS) paths of $k$-th user, when $l=0$, $\alpha_l$, $r_l$ and $\theta_l$ denotes the complex gain of LoS path, distance from $k$-th user respectively, when $l=1,\ldots,P$, $\alpha_l$, $r_l$ and $\theta_l$ denotes the complex gain of $l$-th NLoS path, distance from $l$-th scatter of $k$-th user, respectively.

According to  (\ref{eq3}) and (\ref{eq4}), if the near field is ignored, the perfectly estimated far-field channel can be expressed as 
\begin{equation}
    \label{eq9}
    \mathbf{h}^{far}_k=\alpha_0\mathbf{a}(\theta_0)+\sqrt{\frac{1}{P}}\sum_{l=1}^P\alpha_l\mathbf{a}(\theta_l). 
\end{equation}
In this case, the spectral efficiency is given by \cite{Zhang2021BeamFF}
\begin{equation}
    \label{eq10}  
    R=\sum_{k=1}^K R_{k}=\sum_{k=1}^K \log _{2}\left(1+\frac{\left|\mathbf{h}_{k}^H \mathbf{f}_{k}\right|^{2}}{\sum_{i=1,i\neq k}^K \left|\mathbf{h}_{k}^H \mathbf{f}_{i}\right|^{2}+\sigma_{n}^{2}}\right),
\end{equation}
where $R_k$ denotes the achievable rate of the $k$-th user. Note that (\ref{eq10}) is calculated using the true near-field channel in (\ref{eq8}). 

{\bf Advantages of near field for communication:} As can be seen from the communication model, the near-field channel increases the distance dimension. This reduces the multiuser channel interference providing new DoF.
\

\subsubsection{Sensing Model} For sensing services, it is widely known that the radar transmitting direction map is an important representation of radar performance \cite{4350230}, and it is also an important measure of sensing performance in the ISAC system. The classic directional map (beampattern gain) expresses the transmitted power in a given direction $\theta$ as
\begin{equation}
    \label{eq11}    G(\theta)=\mathbf{a}^H(\theta)\mathbf{R_x}\mathbf{a}(\theta),
\end{equation}
where $\mathbf{a}(\theta)$ is given by $\mathbf{a}^{far}(\theta)$ in (\ref{eq3}), $\mathbf{R}_x$ is the covariance matrix of the transmitted signal. Similarly, in the near field, a directional map expresses the transmitted power in a given point coordinates $(r,\theta)$
\begin{equation}
    \label{eq12} 
    G_n(r,\theta)=\mathbf{a}^H(r,\theta)\mathbf{R_x}\mathbf{a}(r,\theta),
\end{equation}
Based on the settings of (\ref{eq7}), $\bf{R}_x$ in (\ref{eq11}) and (\ref{eq12}) can be written as
\begin{equation}
    \label{eq13}
    \begin{split}
    \mathbf{R}_x&=\mathbb{E}\left\{\mathbf{F}\mathbf{S}\mathbf{S}^H\mathbf{F}^H\right\}\\&=\mathbf{F}\mathbb{E}\left\{\mathbf{S}\mathbf{S}^H\right\}\mathbf{F}^H=\mathbf{F}\mathbf{F}^H.
    \end{split}
\end{equation}
According to (\ref{eq13}), we transform the design covariance matrix in \cite{Liu2017TowardDR} into the design of the transmit beamforming matrix.
In addition, the process of receiving an echo of the near-field target can be modeled as
\begin{equation}
    \label{eq14}
    \mathbf{Y}_R=\beta \mathbf{b}(r_r,\theta_r)\mathbf{a}^H(r_t,\theta_t)\mathbf{X}+\mathbf{Z},
\end{equation}
where $\bf{b}(\cdot)$ is the focusing vector of the receiving antenna array, and has the similar form to $\mathbf{a}^{near}$, $\mathbf{b}(r,\theta)=\big[e^{-j2\pi \frac{r_1-r}{\lambda}},\ldots, e^{-j2\pi \frac{r_{N_r}-r}{\lambda}}\big]^T$; $\beta$ includes reflection coefficient and path loss; $r_r$ and $\theta_r$, $r_t$ and $\theta_t$, represent the distance and angle from the target to the transmitting and receiving arrays, respectively; $\mathbf{X}$ has been denoted by (\ref{eq7}); $\mathbf{Z}\in\mathbb{C}^{N_r \times L}$ is independent, zero-mean complex Gaussian noise matrix with known covariance $\mathbf{R}_w=\sigma_w^2\mathbf{I}_{N_r}$ and $L$ can be interpreted as the number of snapshots of a radar pulse.
\begin{figure}
    \centering
    \includegraphics[width=0.9\linewidth]{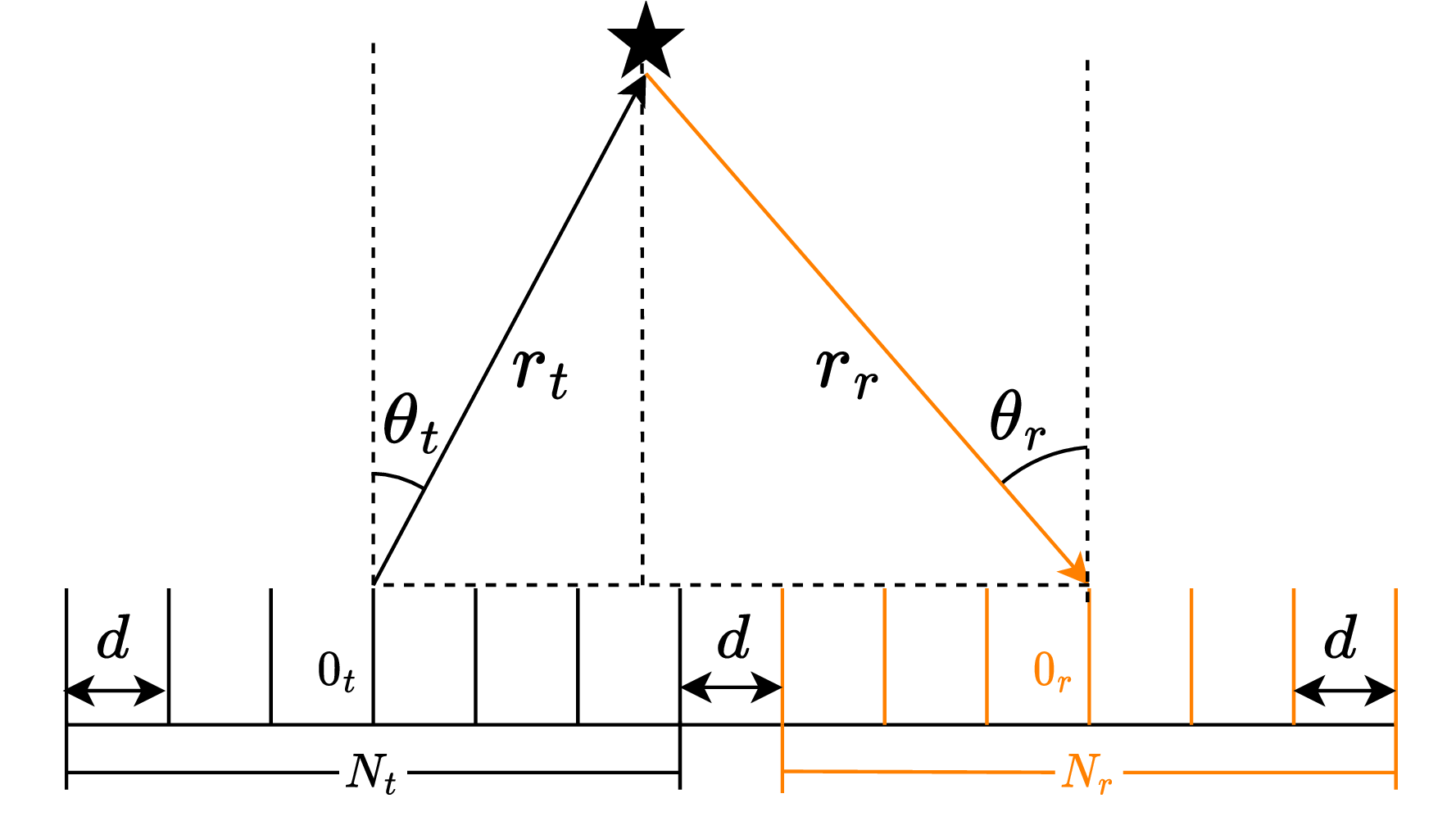}
    \caption{Diagram of the distance and angle from the target to the receiving and transmitting antennas.}
    \label{fig5}
\end{figure}
In addition, as shown in Fig. \ref{fig5}, $r_r$, $\theta_r$, $r_t$ and $\theta_t$, have the following relations
\begin{equation}\label{eq15}
    \begin{split}
    &r_r^2=r_t^2+\left(\frac{(N_t+N_r)d}{2}\right)^2-r_td(N_t+N_r)\sin{\theta_t},\\
    &\sin{\theta_r}=\frac{\frac{(N_t+N_r)d}{2}-r_t\sin{\theta_t}}{r_r}.
    \end{split}
\end{equation}
{\bf Advantages of near field for sensing:} From the radar model, it can be seen that the echoes contain distance information. Thus jointly estimating the angle and distance to the target becomes possible.

\begin{remark}
\rm We do not consider interference between transceiver antenna arrays. In fact, the model shown in Fig. 5 is similar to a bistatic radar. With increasing the distance between the receiving and transmitting arrays, the interference between the transceiver antennas can be eliminated by the principle of bistatic radar.
\end{remark}

\section{Flexible trade-off beamforming design}
As in \cite{Liu2018HybridBW}, the authors aim  to design the beamforming matrix so that it has good sensing and communication performance. Therefore, beamforming matrix $\mathbf{F}$ should approximate the communication optimally using the fully digital precoding matrix $\mathbf{F}_{com}$, while approximating the radar beamforming matrix $\mathbf{F}_{rad}$. 
Since zero-forcing (ZF) beamforming is known to approach the total capacity asymptotically in broadcast channels\cite{1603708}, showing its potential to approach the best performance in settings involving multi-user communications.
Thus, the full-digital precoder is designed by ZF, expressed as
\begin{equation}
    \label{eq16}   
\mathbf{F}_{ZF}=\mathbf{H}^H\left(\mathbf{H} \mathbf{H}^{H}\right)^{-1},
\end{equation}
We designed the $\mathbf{F}_{rad}$ using fixed beamforming algorithms, which can be expressed as
\begin{equation}
    \label{eq17}
    \mathbf{F}_{rad}=\mathbf{a}^*(r_t,\theta_t)\in\mathbb{C}^{N_t \times 1}.
\end{equation}
The resulting beampattern by $\mathbf{F}_{rad}$ is mainly used for tracking targets at interested angles. To meet the transmit power budget, we reformulate (\ref{eq16}) and (\ref{eq17}) as
\begin{equation}
\begin{split}
    \label{18}
     \mathbf{F}_{com}&=\sqrt{P_t}\frac{\mathbf{F}_{ZF}}{\|\mathbf{F}_{ZF}\|_F}, \\ \mathbf{F}_{rad}&=\sqrt{P_t}\frac{\mathbf{a}^*(r_t,\theta_t)}{\|\mathbf{a}^*(r_t,\theta_t)\|}.
\end{split}
\end{equation}
When the number of targets and data streams does not coincide, an auxiliary matrix $\mathbf{F}_u\in\mathbb{C}^{1 \times N_s}$ should be constructed which can be adjusted for dimensionality. Since the unitary matrix does not affect the radar beam pattern generated by $\mathbf{F}_{rad}$, the auxiliary matrix should respect the limits of the unitary matrix. Based on the above analysis, the problem can be formulated as
\
\newline
\begin{equation}\label{eq19}
    \begin{split}
    &\min_{\mathbf{F}, \mathbf{F}_u} \eta\left\|\mathbf{F}-\mathbf{F}_{com}\right\|_{F}^{2}+(1-\eta)\left\|\mathbf{F}-\mathbf{F}_{rad} \mathbf{F}_u\right\|_{F}^{2} \\&~~s.t.~~ \left\|\mathbf{F}\right\|_{F}^{2}=P_{t},\mathbf{F}_u\mathbf{F}_u^{H}=1,
    \end{split}
\end{equation}
\
\newline
where $0 <\eta< 1$ is the weighting factor for weighing communication and sensing functions, and $P_t$ is the total transmitting power of the ISAC system.
\begin{remark}
\rm Note that we encapsulate the near-field properties in (\ref{eq4}) and further encapsulate the near-field communication beam properties in the near-field channel. We aim to compare the effects of model mismatch, so ZF precoding based on different model channels and radar beam design based on different array response vectors are reasonable.
\end{remark}

\subsection{Algorithm for Solving $\mathbf{F}$}
Note that the objective function in (\ref{eq19}) is a combination of two Frobenius norms, which can be transformed into the following form
\begin{normalsize}
\begin{equation}\label{eq22}
    \begin{split}
        &\eta\left\|\mathbf{F}-\mathbf{F}_{com}\right\|_{F}^{2}+(1-\eta)\left\|\mathbf{F}-\mathbf{F}_{rad} \mathbf{F}_u\right\|_{F}^{2}=\\&\left\|\left[\sqrt{\eta}\mathbf{I}_{N_t},\sqrt{1-\eta}\mathbf{I}_{N_t}\right]^T\mathbf{F}-\left[\sqrt{\eta}\mathbf{F}_{com}^T,\sqrt{1-\eta}\mathbf{F}_u^T\mathbf{F}_{rad}^T\right]^T\right\|_F^2.
    \end{split}
\end{equation}
\end{normalsize}
Consider that $\mathbf{A}=\left[\sqrt{\eta}\mathbf{I}_{N_t},\sqrt{1-\eta}\mathbf{I}_{N_t}\right]^T$ and $\mathbf{B}=\left[\sqrt{\eta}\mathbf{F}_{com}^T,\sqrt{1-\eta}\mathbf{F}_u^T\mathbf{F}_{rad}^T\right]^T$, then (\ref{eq19}) can be simplified as
\begin{equation}
    \label{eq23}
    \begin{split}
        &\min_{\mathbf{F}} \left\|\mathbf{A}\mathbf{F}-\mathbf{B}\right\|_{F}^{2} \\&s.t.~ \left\|\mathbf{F}\right\|_{F}^{2}=P_{t},
    \end{split}
\end{equation}
which is a quadratically constrained quadratic program (QCQP) problem and can be solved using SDR \cite{2010Semidefinite}. Nevertheless, SDR has a large amount of computation; therefore, we adopt the constrained least square (LS) method with a lower complexity \cite{Guo2022NonuniformBP}. The objective function in (\ref{eq23}) is a LS problem and it has a solution $\mathbf{F}=\mathbf{A}^{\dagger}\mathbf{B}$. To satisfy the power constraint, we multiply the LS solution by a constraint factor $\frac{\sqrt{P}_t}{\|\mathbf{F}\|_{F}}$ as the sub-optimal solution of problem ($\ref{eq23}$)
\begin{equation}
    \label{eq24}
    \mathbf{F}=\frac{\sqrt{P}_t}{\|\mathbf{F}\|_{F}}\mathbf{A}^{\dagger}\mathbf{B}.
\end{equation}

\subsection{Obtaining the Auxiliary Matrix $\mathbf{F}_u$}
We reduce the Euclidean distance between $\mathbf{F}_{com}$ and $\mathbf{F}_{rad} $ without affecting communication and radar performance, thus improving the trade-off performance of the integrated system, and optimization problem for the auxiliary matrix $\mathbf{F}_u$ is formulated as
\begin{equation}
    \label{eq20}
    \min_{\mathbf{F}_u} \left\|\mathbf{F}_{com}-\mathbf{F}_{rad}\mathbf{F}_u\right\|_{F}^2\quad s.t.~~\mathbf{F}_u\mathbf{F}_u^{H}=1.
\end{equation}
Problem (\ref{eq20}) is obviously the least squares problem defined over the Stiefel manifold, which is non-convex. Fortunately, \cite{Viklands2006AlgorithmsFT} has shown that (8) can be regarded as an orthogonal procrustes problem (OPP) and has a global closed-form solution based on the SVD, which can be written as
\begin{equation}
    \label{eq21}
    \mathbf{F}_u=\mathbf{U}_1\mathbf{I}_{1 \times N_s}\mathbf{V}_1^H,
\end{equation}
where $\mathbf{U}_1\Sigma_1\mathbf{V}_1^H=\mathbf{F}_{rad}\mathbf{F}_{com}$ is the SVD of $\mathbf{F}_{rad}^H\mathbf{F}_{com}$. $\mathbf{I}_{1 \times N_s}$ consists of a $1 \times 1$ identity matrix and a $1 \times (N_s-1)$ zero matrix.
We summarize Subections III-A and B in Algorithm 1.
\renewcommand{\algorithmicrequire}{\textbf{Input:}}
\renewcommand{\algorithmicensure}{\textbf{Output:}}
\begin{algorithm}[t]
\caption{Proposed Algorithm for solving (\ref{eq19})}
\label{Algorithm1}
\begin{algorithmic}
\Require ~$\mathbf{H},~\mathbf{F}_{rad}, P_t$, weighting factor $0\le \eta\le 1$
\vspace{1mm}
\Ensure ~$\mathbf{F}=\frac{\sqrt{P}_t}{\|\mathbf{F}\|_{F}}\mathbf{A}^{\dagger}\mathbf{B}$
\vspace{1mm}
\Statex{{1.} Compute $\mathbf{F}_{com}$ by (\ref{eq16}).}
\Statex{{2.} Solve  problem (\ref{eq20}) by (\ref{eq21}).}
\Statex{{3.} Compute ${\mathbf{A}}, {\mathbf{B}}$, simplify (\ref{eq19}) as (\ref{eq23}).}
\Statex{{4.} Utilize LS method to solve (\ref{eq23}).}
\Statex{{5.} Multiply the LS solution by a constraint factor $\frac{\sqrt{P}_t}{\|\mathbf{F}\|_{F}}$.}
\end{algorithmic}
\end{algorithm}

\subsection{Alternating Minimization (AM) Algorithm for Solving ($\ref{eq19}$ )\cite{Liu2018HybridBW}}
The idea of \cite{Liu2018HybridBW} is to divide the problem into two sub-problems, one of which is similar to (\ref{eq20}) 
\begin{equation}
    \label{eq25}
    \min_{\mathbf{F}_u} \left\|\mathbf{F}-\mathbf{F}_{rad}\mathbf{F}_u\right\|_{F}^2\quad s.t.~~\mathbf{F}_u\mathbf{F}_u^{H}=1,
\end{equation}
while the other is (\ref{eq23}). The solutions to both subproblems are given by (\ref{eq21}) and (\ref{eq24}). Then, for simplicity, we  can describe the AM algorithm as Algorithm \ref{Algorithm2}.

\begin{algorithm}[t]
\caption{AM algorithm for solving ($\ref{eq19}$) in\cite{Liu2018HybridBW}}
\label{Algorithm2}
\begin{algorithmic}
\Require $\mathbf{H}, \mathbf{F}_{\text {com }}, \mathbf{F}_{\text {rad }}, 0 \leq \eta \leq 1$, $P_{t}$ , tolerable accuracy  $\varepsilon=10^{-6}$ and the maximum iteration number $k_{\max }$ 
\vspace{0.5mm}
\Ensure ~$\mathbf{F},~\mathbf{F}_u$
\vspace{1mm}
\Statex{{1.} Initialize randomly  $\mathbf{F}_{u}^{(0)}$ i.e. [1,1], get $\mathbf{F}$ by (\ref{eq24}), Compute the
objective function of (\ref{eq19}), denoted as $f^{(0)}$. Set k = 1.}
\vspace{1mm}
\Statex{\textbf{while} $k<k_{max}$ or $\left|f^{(k)}-f^{(k-1)}\right|>\varepsilon$}
\vspace{1mm}
\Statex{{2.} Compute $\mathbf{F}_{u}^{(k)}$ by solving (\ref{eq25}).}
\Statex{{3.} Compute $\mathbf{F}^{(k)}$ by solving (\ref{eq23}).}
\Statex{{4.} Compute the objective function $f^{(k)}$ based on the obtained variables.}
\Statex{{5.} $k=k+1$.}
\Statex{\textbf{end While}}
\end{algorithmic}
\end{algorithm}

\section{Parameter Estimation and Cramér–Rao Bound}
In this section, we introduce the 2-D MUSIC estimation algorithm and calculate CRB to compare the optimal parameter estimation performance. 
\subsection{2-D MUSIC Algorithm}
2-D MUSIC is a classical signal estimation algorithm, which can obtain the distance and angle information of signal source using the property that the signal subspace is orthogonal to the noise subspace \cite{Huang1991NearfieldMS}. The signal subspace consists of the eigenvectors corresponding to the first $M$ largest eigenvalues of $\mathbf{R}_{Y}$, where $\mathbf{R}_{Y}$ is the covariance matrix of the radar received signal and $M$ is the number of signal sources. The noise subspace consists of the rest of the eigenvectors. Hence, $\mathbf{R}_{Y}$ can be calculated by exploiting $L$ snapshots
\begin{equation}
    \label{eq4-1}
    \mathbf{R}_Y=\frac{1}{L}\mathbf{Y}_{R}\mathbf{Y}_{R}^H.
\end{equation}
Note that we assume a single target case, i.e., $M=1$, and when the target number is unknown, the target number can be estimated using algorithms such as minimum description length (MDL)\cite{mdl}.
After conducting eigenvalue decomposition
\begin{equation}
    \label{eq4-2}  \mathbf{R}_Y=\mathbf{Q}\Lambda\mathbf{Q}^T,
\end{equation}
where $\Lambda$ is a diagonal matrix whose elements on the diagonal are the eigenvalues in descending order, $\mathbf{Q}$ is an $N \times N$ square matrix with its columns corresponding to eigenvectors. Let $\mathbf{q}_{M+1}, \ldots, \mathbf{q}_{N_r}$ be the noise eigenvectors to construct the noise eigenvector matrix as
\begin{equation}
\label{4-3}
\mathbf{Q}_N=\left[\mathbf{q}_{M+1}, \ldots, \mathbf{q}_{N_r}\right].
\end{equation}
Then, the 2-D MUSIC spectrum function of near-field sources can be presented by 
\begin{equation}
    \label{4-4}
    f(r,\theta)=\frac{1}{\mathbf{b}^H(r,\theta)\mathbf{Q}_N\mathbf{Q}_N^H\mathbf{b}(r,\theta)},
\end{equation}
where $\mathbf{b(r,\theta)}$ is the receiving focusing vector. By performing a two-dimensional spectrum search, matching angles, and distances $\hat{r},\hat{\theta}$ can be obtained as
\begin{equation}
   (\hat{r_r},\hat{\theta_r})={\arg \underset{(r,\theta)}{\max}}\, f(r,\theta).
\end{equation}
Although the parameters obtained are relative to the receiving antenna, it can be transformed to $(\hat{r_t},\hat{\theta_t})$ due to the precise relationship in (\ref{eq15}).
\subsection{Cramér–Rao Bound}
We perform an equivalent transformation of (\ref{eq14}), 
\begin{equation}
    \label{eq33}
    \begin{split} \mathbf{Y}_R=\beta\mathbf{G}\mathbf{X}+\mathbf{W},
    \end{split}
\end{equation}
where $\mathbf{G}=\mathbf{b}(r_r,\theta_r)\mathbf{a}^H(r_t,\theta_t)$.
Then, we vectorize the matrix $\mathbf{Y}_R$
\begin{equation}
    \label{eq34}
\mathbf{y}_r=\mathrm{vec}\left(\mathbf{Y}_R\right)=\mathbf{u}+\mathbf{w},
\end{equation}
where $\mathbf{u}=\mathrm{vec}(\beta\mathbf{G}\mathbf{X})$ and $\mathbf{w}=\mathrm{vec}(\mathbf{W})$.
Let $\boldsymbol{\xi}=[\boldsymbol{\phi} ,\tilde{\boldsymbol{\beta}}]$, where $\boldsymbol{\phi}=\left[r_t,\theta_t\right]^T$ and $\tilde{\boldsymbol{\beta}}=\left[\operatorname{Re}(\beta), \operatorname{Im}(\beta)  \right]^T$  denotes the unknown parameters to estimate. According to the assumptions of zero-mean complex Gaussian noise, complex Gaussian vector  can be written as $\mathbf{y}_r\sim\mathcal{CN} \big(\mathbf{u}, \sigma_w^2\mathbf{I}\big)$. Then, the Fisher information matrix (FIM) for
estimating $\pmb{\xi}$ from $\mathbf{y}_r$ is given by \cite{Kay1993FundamentalsOS,Bekkerman2006TargetDA}
\begin{equation}
    \label{eq35}    \left[\mathbf{J}_{\boldsymbol{\xi}}\right]_{i, j}=\frac{2}{\sigma_{w}^{2}} \operatorname{Re}\left\{\frac{\partial \mathbf{u}^{H}}{\partial \boldsymbol{\xi}_{i}} \frac{\partial \mathbf{u}}{\partial \boldsymbol{\xi}_{j}}\right\},\ i,j=1,\ldots,4.
\end{equation}
As a consequence, $\mathbf{J}_{\boldsymbol{\xi}}$ can be partitioned as
\begin{equation}
    \label{eq36}
\mathbf{J}_{\boldsymbol{\xi}}=\left[\begin{array}{ll}
\mathbf{J}_{\boldsymbol{\phi} \boldsymbol{\phi}} & \mathbf{J}_{\boldsymbol{\phi} \tilde{\boldsymbol{\beta}}} \\
\mathbf{J}_{\boldsymbol{\phi} \tilde{\boldsymbol{\beta}}}^{T} & \mathbf{J}_{\tilde{\boldsymbol{\beta}} \tilde{\boldsymbol{\beta}}}
\end{array}\right].
\end{equation}
According to  \cite{Kay1993FundamentalsOS}, CRB is given by the inverse of the FIM. Then, the CRB of $r_t$ and $\theta_t$ can be expressed as
\begin{subequations}
    \label{eq37}
    \begin{align}
    \operatorname{CRB}(r_t)&=\left[\operatorname{CRB}(\boldsymbol{\phi})\right]_{1,1},\\
    \operatorname{CRB}(\theta_t)&=\left[\operatorname{CRB}(\boldsymbol{\phi})\right]_{2,2},\\
    \operatorname{CRB}(\boldsymbol{\phi})&=\left(\mathbf{J}_{\boldsymbol{\phi} \boldsymbol{\phi}}-\mathbf{J}_{\boldsymbol{\phi} \tilde{\boldsymbol{\beta}}} \mathbf{J}_{\tilde{\boldsymbol{\beta}} \tilde{\boldsymbol{\beta}}}^{-1} \mathbf{J}_{\boldsymbol{\phi} \tilde{\boldsymbol{\beta}}}^{T}\right)^{-1},
    \end{align}
\end{subequations}
where the expressions of the matrices $\mathbf{J}_{\boldsymbol{\phi} \tilde{\boldsymbol{\beta}}}$, $\mathbf{J}_{\boldsymbol{\phi} \tilde{\boldsymbol{\phi}}}$ and $\mathbf{J}_{\boldsymbol{\beta} \tilde{\boldsymbol{\beta}}}$ are shown below and derived in \appref{A2}.
\begin{align}
    J_{{\phi}_l {\phi}_p}&=\frac{2|\beta|^2L}{\sigma_{w}^{2}} \operatorname{Re}\left\{ \operatorname{tr}\left(  \dot{\mathbf{G}}_{\phi_{p}}\mathbf{R}_x\dot{\mathbf{G}}_{\phi_{l}}^H\right)\right\}, l,p\in\{1,2\}, \label{eq38}\\ 
    \mathbf{J}_{\boldsymbol{\phi} \tilde{\boldsymbol{\beta}}}
    &=\frac{2L}{\sigma_{w}^{2}} \operatorname{Re}\left\{\left[\begin{array}{ll} 
    \beta^{*} \operatorname{tr}\left(\mathbf{G}\mathbf{R}_x\dot{\mathbf{G}}_{\phi_1}^H\right)\\
    \beta^{*} \operatorname{tr}\left(\mathbf{G}\mathbf{R}_x\dot{\mathbf{G}}_{\phi_2}^H\right)\end{array}\right]
    \left[1, j\right]
    \right\},\label{eq39} \\
    \mathbf{J}_{\tilde{\boldsymbol{\beta}} \tilde{\boldsymbol{\beta}}} 
    &=\frac{2L}{\sigma_{w}^{2}}\mathbf{I}_2 \operatorname{tr}\left(\mathbf{G}\mathbf{R}_x\mathbf{G}^H\right),\label{eq40}
\end{align}
where $\dot{\mathbf{G}}_{\phi_1}=\frac{\partial\mathbf{G}}{\partial r_t},~\dot{\mathbf{G}}_{\phi_2}=\frac{\partial \mathbf{G}}{\partial \theta_t}$. By bringing (\ref{eq38})-(\ref{eq40}) into (\ref{eq37}),   (\ref{CRB}) is obtained\footnote{In order to solve the situation that no inverse matrix exists for the FIM matrix when deriving (\ref{eq42}) and (\ref{eq43}), here we replace the original inverse operation with a pseudo-inverse.}, where $\operatorname{SNR}_{r}=\frac{|\beta|^2LP_t}{\sigma_{w}^{2}}$ represents the radar receive SNR of the echo signal. 

\begin{figure*}[!t]
\normalsize
\setcounter{MYtempeqncnt}{\value{equation}}
\setcounter{equation}{40}

\begin{footnotesize}
\begin{equation}
\begin{split}
\label{CRB}
&\normalsize{\operatorname{CRB}(\boldsymbol{\phi})=}\frac{P_t\operatorname{tr}\left(\mathbf{G}\mathbf{R}_x\mathbf{G}^H\right)}{2\operatorname{SNR}_{r}}\cdot\\ 
     &\left[\setlength{\arraycolsep}{1.2pt} \begin{array}{ll}
      \operatorname{tr}\left(\dot{\mathbf{G}}_{\phi_{1}}\mathbf{R}_x\dot{\mathbf{G}}_{\phi_{1}}^H\right)\operatorname{tr}\left(\mathbf{G}\mathbf{R}_x\mathbf{G}^H\right)-\left|\operatorname{tr}\left(\mathbf{G}\mathbf{R}_x\dot{\mathbf{G}}_{\phi_1}^H\right)\right|^2 \
      &\operatorname{Re}\left\{\operatorname{tr}\left(\dot{\mathbf{G}}_{\phi_{2}}\mathbf{R}_x\dot{\mathbf{G}}_{\phi_{1}}^H\right)\operatorname{tr}\left(\mathbf{G}\mathbf{R}_x\mathbf{G}^H\right)-\operatorname{tr}\left(\mathbf{G}\mathbf{R}_x\dot{\mathbf{G}}_{\phi_1}^H\right)\operatorname{tr}\left(\mathbf{G}\mathbf{R}_x\dot{\mathbf{G}}_{\phi_2}^H\right)^{*}\right\}\\\\
      \operatorname{Re}\left\{\operatorname{tr}\left(\dot{\mathbf{G}}_{\phi_{1}}\mathbf{R}_x\dot{\mathbf{G}}_{\phi_{2}}^H\right)\operatorname{tr}\left(\mathbf{G}\mathbf{R}_x\mathbf{G}^H\right)-\operatorname{tr}\left(\mathbf{G}\mathbf{R}_x\dot{\mathbf{G}}_{\phi_1}^H\right)\operatorname{tr}\left(\mathbf{G}\mathbf{R}_x\dot{\mathbf{G}}_{\phi_2}^H\right)^{*}\right\}\
      &\operatorname{tr}\left(\dot{\mathbf{G}}_{\phi_{2}}\mathbf{R}_x\dot{\mathbf{G}}_{\phi_{2}}^H\right)\operatorname{tr}\left(\mathbf{G}\mathbf{R}_x\mathbf{G}^H\right)-\left|\operatorname{tr}\left(\mathbf{G}\mathbf{R}_x\dot{\mathbf{G}_{\phi_2}}^H\right)\right|^2
 \end{array}\right]^{\dag}
\end{split}
\end{equation}
\end{footnotesize}
\setcounter{equation}{\value{MYtempeqncnt}}
\hrulefill
\vspace*{4pt}
\end{figure*}

Specifically, when the distance to the target is known, i.e. when $r_t$ is a constant, we can derive that
\
\\
\setcounter{equation}{41}
\begin{small}
\begin{equation}
    \begin{split}
    \label{eq42}
    &\operatorname{CRB}(\theta_t)=\\&\frac{P_t\operatorname{tr}\left(\mathbf{G}\mathbf{R}_x\mathbf{G}^H\right)}{2\operatorname{SNR}_{r}\left(\operatorname{tr}\left(\dot{\mathbf{G}}_{\phi_{2}}\mathbf{R}_x\dot{\mathbf{G}}_{\phi_{2}}^H\right)\operatorname{tr}\left(\mathbf{G}\mathbf{R}_x\mathbf{G}^H\right)-\left|\operatorname{tr}\left(\mathbf{G}\mathbf{R}_x\dot{\mathbf{G}_{\phi_2}}^H\right)\right|^2\right)},
    \end{split}
\end{equation}
\end{small}
when the angle of the target is known, i.e., $\theta_t$ is a constant, 
\begin{small}
\begin{equation}
    \begin{split}
    \label{eq43}
    &\operatorname{CRB}(r_t)=\\&\frac{P_t\operatorname{tr}\left(\mathbf{G}\mathbf{R}_x\mathbf{G}^H\right)}{2\operatorname{SNR}_{r}\left(\operatorname{tr}\left(\dot{\mathbf{G}}_{\phi_{1}}\mathbf{R}_x\dot{\mathbf{G}}_{\phi_{1}}^H\right)\operatorname{tr}\left(\mathbf{G}\mathbf{R}_x\mathbf{G}^H\right)-\left|\operatorname{tr}\left(\mathbf{G}\mathbf{R}_x\dot{\mathbf{G}_{\phi_1}}^H\right)\right|^2\right)},
    \end{split}
\end{equation}
\end{small}
which is the same as the CRB of the single target angle estimate expressed in \cite{Bekkerman2006TargetDA}. The derivation of equations (41)-(43) is given in the \appref{A3}.

\section{Power-saving beamforming design}
Section III compares the performance of NFBF and FFBF with fixed power constraints. In this section, we specifically compare power consumption when meeting communication and perceived QoS requirements.

We choose the received SINR of the user as the communication performance, then the SINR of $k$-th user can be expressed as
\begin{equation}
\begin{split}
    \label{eq5-1}
    \gamma_{k}&=\frac{\left|\mathbf{h}_{k} \mathbf{f}_{k}\right|^{2}}{\sum_{i=1, i \neq k}^{K}\left|\mathbf{h}_{k} \mathbf{f}_{i}\right|^{2}+\sigma_{n}^{2}}\\&=\frac{\operatorname{tr}\left(\mathbf{h}_{k}\mathbf{F}_{k}\mathbf{h}_{k}^{H} \right)}{\operatorname{tr}\left(\mathbf{h}_{k}\sum_{\substack{i=1,i \neq k}}^{K} \mathbf{F}_{i} \mathbf{h}_{k}^{H} \right)+\sigma_{n}^{2}},
    \end{split}
\end{equation}
where $\mathbf{F}_k=\mathbf{f}_k\mathbf{f}_k^H$. Note that the covariance matrix of (\ref{eq13}) becomes equivalent to
\begin{equation}
    \label{5-2}
    \mathbf{R}_x=\sum_{k=1}^{K} \mathbf{F}_{k},
\end{equation}
and the transmit power is given by
\begin{equation}
    \label{5-3} \left\|\mathbf{F}\right\|_F^2=\operatorname{tr}(\mathbf{R}_x)=\sum_{k=1}^{K}\operatorname{tr}\left(\mathbf{F}_{k}\right).
\end{equation}
We choose CRB on the estimated performance as a sensing performance constraint and formulate the problem as
\begin{subequations}\label{5-4}
\begin{align}
       \min _{\mathbf{F}} ~&\sum_{k=1}^{K} \operatorname{tr}\left( \mathbf{F}_{k}\right) \label{5-2a}\\
    \text { s.t. } 
    ~& \gamma_{k}\ge\Gamma_k,~ \forall k,\\
    &\operatorname{CRB}\left({\theta_t}\right)<\varepsilon_1,\operatorname{CRB}\left(r_t\right)<\varepsilon_2\label{5-2b}\\
    & \mathbf{F}_{k} \succeq 0, \mathbf{F}_{k}=\mathbf{F}_{k}^{H}, \operatorname{rank}\left(\mathbf{F}_{k}\right)=1, \quad \forall k.
\end{align}
\end{subequations}
where $\Gamma_k$ is the threshold of SINR for $k$-th user. Then without loss of generality, from the analysis of (\ref{eq42}) and (\ref{eq43}), there is a negative correlation between $\operatorname{tr}\left(\mathbf{G}\mathbf{R}_x\mathbf{G}^H\right)$ 
 and the value of $\operatorname{CRB}(r_t)$ or $\operatorname{CRB}(\theta_t)$. 
\begin{equation}
     \label{5-5}
\operatorname{tr}\left(\mathbf{G}\mathbf{R}_x\mathbf{G}^H\right)=\operatorname{tr}\left(\mathbf{b}\mathbf{a}^H\mathbf{R}_x\mathbf{a}\mathbf{b}^H\right)=N_rG_n(r_t,\theta_t),
\end{equation}
where $\mathbf{b}$, $\mathbf{a}$ denote $\mathbf{b}(r_r,\theta_r)$, $\mathbf{a}(r_t,\theta_t)$, $G_n$ is the transmitted power at the target. Thus, the constraint in (\ref{5-2b}) can be approximated using a constraint on $G_n$. Hence, (\ref{5-2}) can be re-written as
\begin{equation}
    \label{eq5-6}
    \begin{split}
    \min _{\mathbf{F}} ~&\sum_{k=1}^{K} \operatorname{tr}\left( \mathbf{F}_{k}\right) \\
    \text { s.t. } ~&\gamma_{k}\ge\Gamma_k,~ \forall k,\\
    &G_n(r_t,\theta_t)\ge \hat{G},\\
    & \mathbf{F}_{k} \succeq 0, \mathbf{F}_{k}=\mathbf{F}_{k}^{H}, \operatorname{rank}\left(\mathbf{F}_{k}\right)=1, \quad \forall k.
    \end{split}
\end{equation}
It is clear that (\ref{eq5-6}) is non-convex due to the rank-1 constraint. Nonetheless, a suboptimal solution can be obtained by the SDR technique. By neglecting  the rank-1 constraints, (\ref{eq5-6}) becomes a standard semidefinite program (SDP), which can be efficiently solved via numerical solvers, e.g., the CVX toolbox.  For nonrank-1 solutions, an approximated solution is obtained by standard rank-1 approximation techniques, such as  Gaussian randomization \cite{2010Semidefinite}. Then, $\mathbf{f}_k$ can be recovered from $\mathbf{F}_k$.

\section{Simulations and Discussions}
In this section, we present the simulation results that validate the effectiveness of the proposed NFBF design for ISAC systems with XL-array. Our simulation setup assumes an ISAC BS that has $N_t=N_r=256$ antennas uniformly distributed in lines at half-wavelength spacing and the carrier frequency is set to 30 GHz. The simulation parameters are set as follows: we consider a scenario with $K=2$ users and $N_s=2$ data streams, and the total power is set to $P_t=30$ dBm. Assuming that the users are located at the same angle at different distances, the locations are set as $[5 \rm{~m}, 0^{\circ}]$ and $[15\rm{~m}, 0^{\circ}]$, with each user having $P=2$ scattering paths in their channel. We assume that the target is located at $[5\rm{~m}, 60^{\circ}]$, and the communication symbol or radar pulse length is set as $L=64$. Note that the angle of the user or target is the angle with the normal of the transmitting antenna array, like the one shown in Fig. \ref{fig5}. The above settings remain unchanged in the simulation unless otherwise specified.

\begin{figure}[!h]
  \centering
   \subfigure[Angle estimation.]{\label{crb1}	\includegraphics[width=1\linewidth]{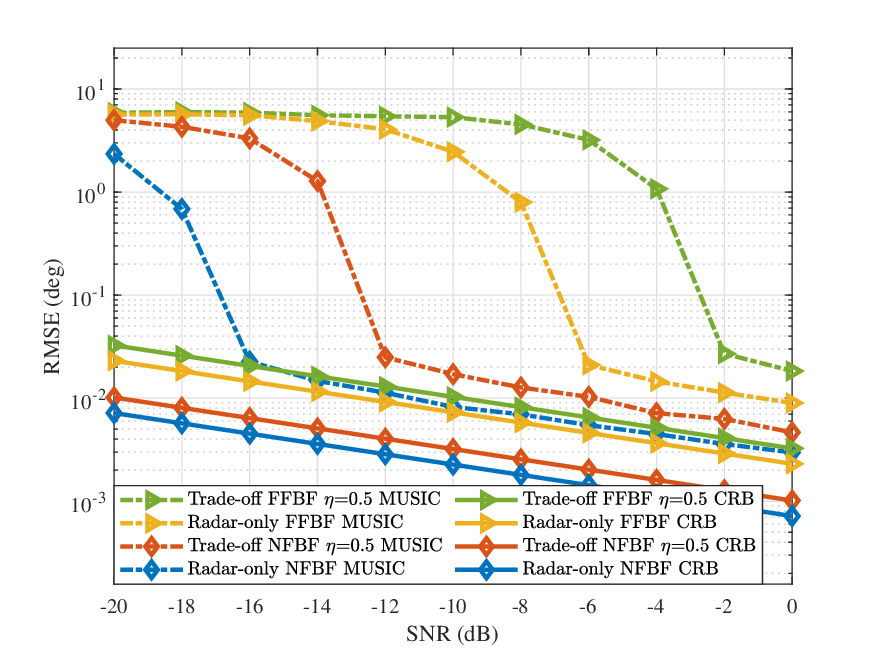}}
   \subfigure[Distance estimation.]{\label{crb2}	\includegraphics[width=1\linewidth]{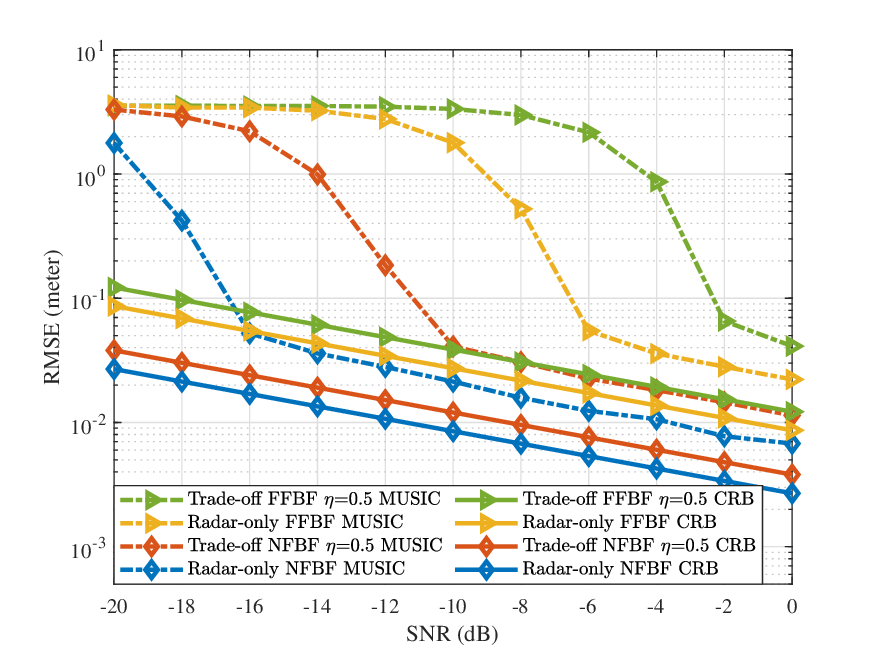}}
  
  \caption{Parameters estimation performance for different beam designs at different SNR.}
\end{figure}

\begin{figure}[t]
  \centering
  \includegraphics[width=1\linewidth]{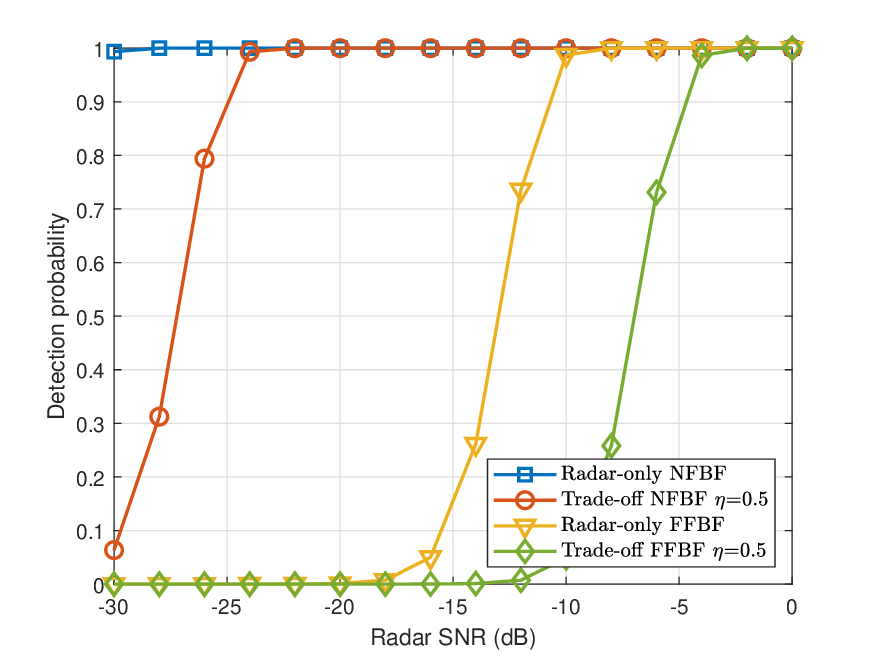}
  \caption{ Detection probability comparisons among different approaches.}
  \label{Pd}
\end{figure}
\begin{figure}[t]
  \centering
  \includegraphics[width=1\linewidth]{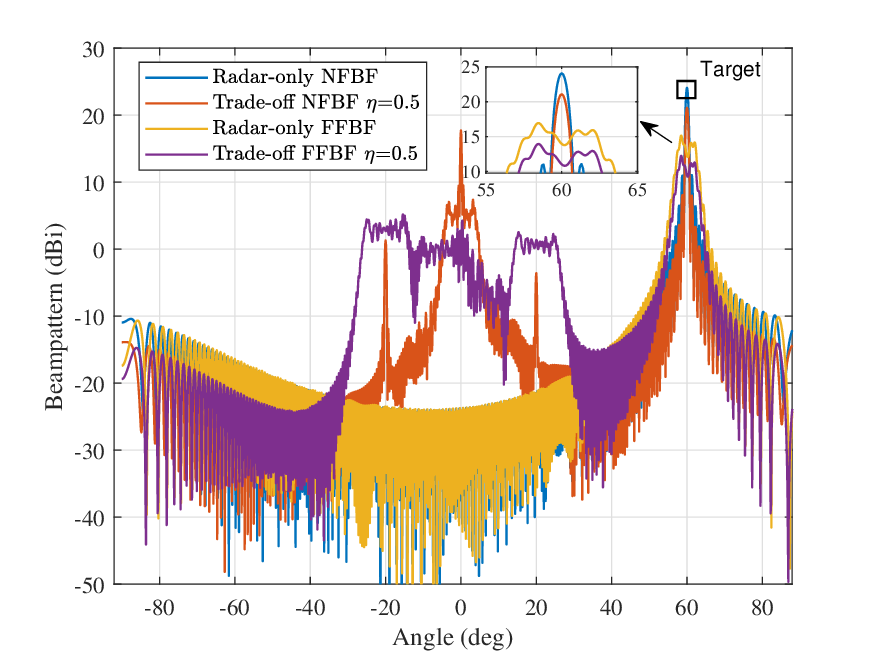}
  \caption{Radar beampatterns at $5\rm{~m}$ from the BS obtained by different approaches.}
  \label{pattern}
\end{figure}
\subsection{Parameters Estimation}
We first show the performance of parameter estimation, including angle and distance for beams designed by different methods in Fig. \ref{crb1} and Fig. \ref{crb2}, where the trade-off factor is set as $\eta=0.5$ and 'Radar-only' denotes the ideal radar beamformer. All graphs of CRB here are plotted using the closed-form solution derived in (\ref{CRB}). We also employ 500 Monte-Carlo trials using 2-D MUSIC to validate the performance of CRB and define the root mean squared error (RMSE) as 
\begin{equation}
    \operatorname{RMSE}=\sqrt{\frac{1}{500} \sum_{n=1}^{500}\left(\hat{\theta_t}_{(n)}-\theta_t\right)^{2}}.
\end{equation}
In our evaluation of the proposed NFBF designs, we analyze the performance of the 2-D MUSIC algorithm in estimating the target parameters. Fig. \ref{crb1} and Fig. \ref{crb2} show that the RMSE estimated by the 2-D MUSIC algorithm can close to the CRB. 
This indicates that the CRB is valid, and the proposed NFBF can improve the parameter estimation performance. To further reinforce this conclusion, we plot the detection probability curve in Fig. \ref{Pd}, where the false alarm probability is set to $10^{-7}$. The results confirm that the proposed NFBF designs significantly improve the detection probability compared to FFBF approach.

\subsection{Sensing Performance}
To demonstrate the effectiveness of our proposed NFBF designs, we present the radar beampatterns at 5 m from the BS in Fig. \ref{pattern}. We compare the trade-off design with a weighting factor of 0.5 and the radar-only design. As shown in Fig. \ref{pattern}, our proposed NFBF design has higher power at the target location than FFBF design. This is one of the key reasons why the beam of the NFBF design has better estimated performance, as evidenced by the lower CRB \cite{Liu2021CramrRaoBO}.

\begin{figure}[!h]
  \centering
  \includegraphics[width=1\linewidth]{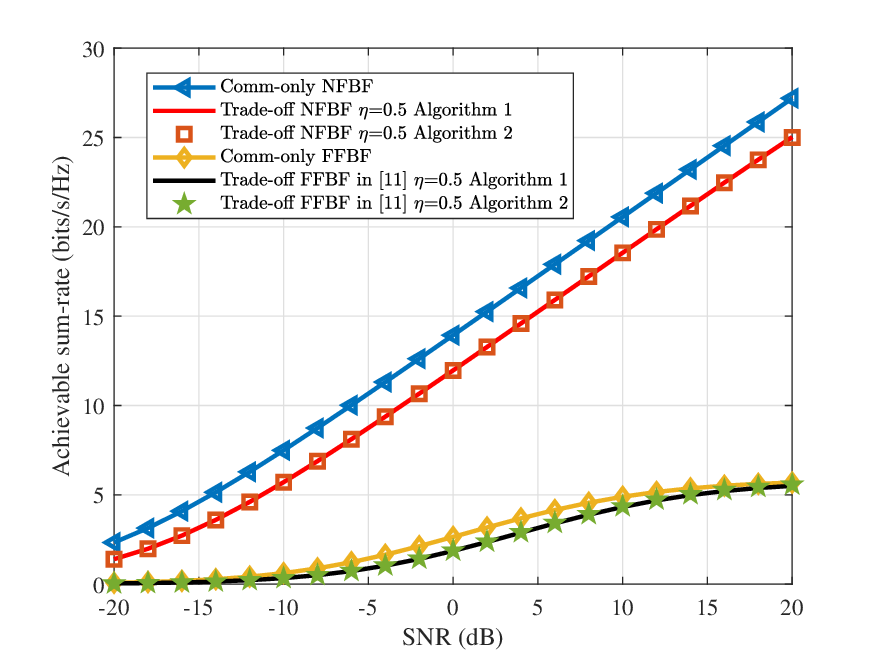}
  \caption{Achievable sum-rate for different approaches versus transmit SNR.}
  \label{sumrate}
\end{figure}

\begin{figure}[h]
  \centering
  \subfigure[User1.]{{\label{f1}}\includegraphics[width=1\linewidth]{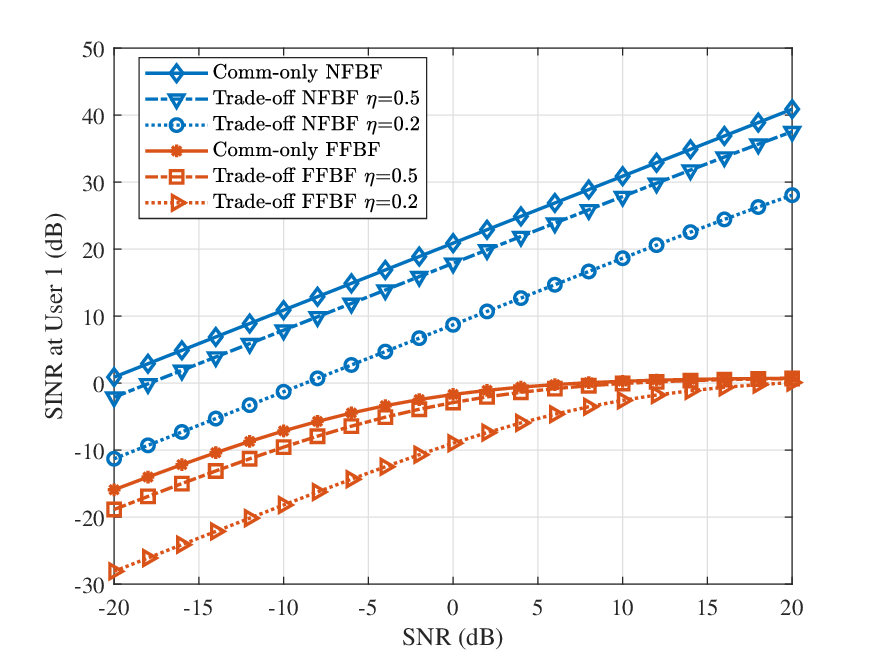}}
  \subfigure[User2.]{{\label{f2}}\includegraphics[width=1\linewidth]{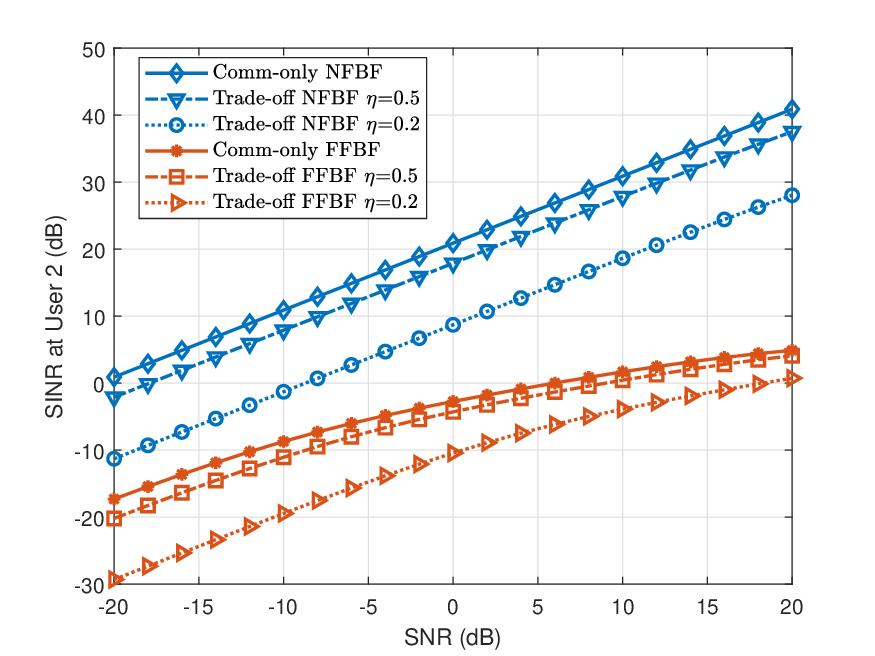}}
  \caption{The received SINR at each user of different approaches versus transmit SNR.}
  \label{F}
\end{figure}

\subsection{Communication Performance}
To evaluate the communication performance of the proposed beamforming design, we conduct simulations to observe the communication rate profile. In Fig. \ref{sumrate}, we compare the theoretical communication rates of beams designed by different approaches, where $\eta=0.5$, and `Comm-only' denotes the upper bound on the communication rate obtained from the ZF-based fully-digital communication beamformer. 
The results demonstrate that the proposed NFBF design effectively improves achievable sum-rate. And the reachable sum rate of the FFBF design reaches an upper limit with increasing SNR. That is because NFBF can manage interference between users located at the same angle very well, while FFBF cannot. With increasing SNR under the transmit power constraint, the FFBF design makes the interference power much larger than the noise power, which makes the achievable sumrate saturate at higher SNRs. And due to the mismatch phenomenon illustrated in Lemma 1, the signal power received by the FFBF users is greatly reduced, resulting in the very low sum rate upper limit.

Furthermore, we compare two algorithms for solving (\ref{eq19}). We find that both algorithms achieve the same beam performance, but Algorithm 1 does not require multiple iterations and is faster to solve.


Moreover, Fig. \ref{F} illustrates the received SINR at each user versus transmit SNR. In the high SNR region in Fig. \ref{F}, the curve corresponding to 'Comm-only FFBF' saturates, which is consistent with the analysis in Fig. \ref{sumrate}. Since the weighting factor changes the power of the communication beam, it converges to 'Comm-only FFBF' at higher SNRs. In contrast, NFBF can focus energy and manage inter-user interference at the same angle through ZF, so that the user received SINR can increase with increasing SNR.

\begin{figure}[t]
  \centering
  \includegraphics[width=1\linewidth]{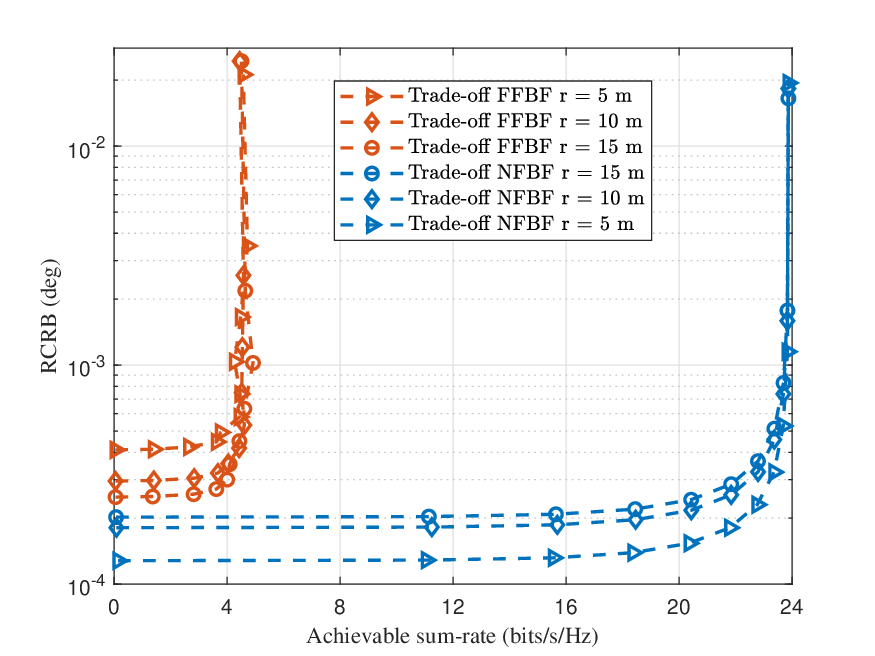}
  \caption{RCRB versus communication rate, a single target is located at different distance $r$ and SNR= 15dB.}
  \label{rate eta}
\end{figure}

\subsection{Trade-off Performance between Communication and Sensing}
Next, we analyze the trade-off performance between communication and sensing under the NFBF and FFBF designs in Fig. 11, where the sensing performance is evaluated via the root Cramer-Rao bound (RCRB). The curve clearly illustrates the existence of a trade-off, regardless of whether NFBF or FFBF is employed.
 The communication rate of FFBF saturates around $5$ bit/Hz/s quickly, whereas NFBF reaches a saturation rate of approximately $24$ bit/Hz/s. Furthermore, it can be found that the estimation performance of FFBF deteriorates with a decrease in target distance, which contrasts with NFBF. The degradation is attributed to the more substantial gain loss experienced by FFBF in the near field as the distance decreases, resulting in lower power at the target and poorer estimation performance. Overall, the analysis indicates that NFBF exhibits a better trade-off performance in this regard.

\subsection{Effect of Target Distance on Sensing}
\begin{figure}[t]
  \centering
  \subfigure[Distance estimation.]  {{\label{crb22}}\includegraphics[width=1\linewidth]{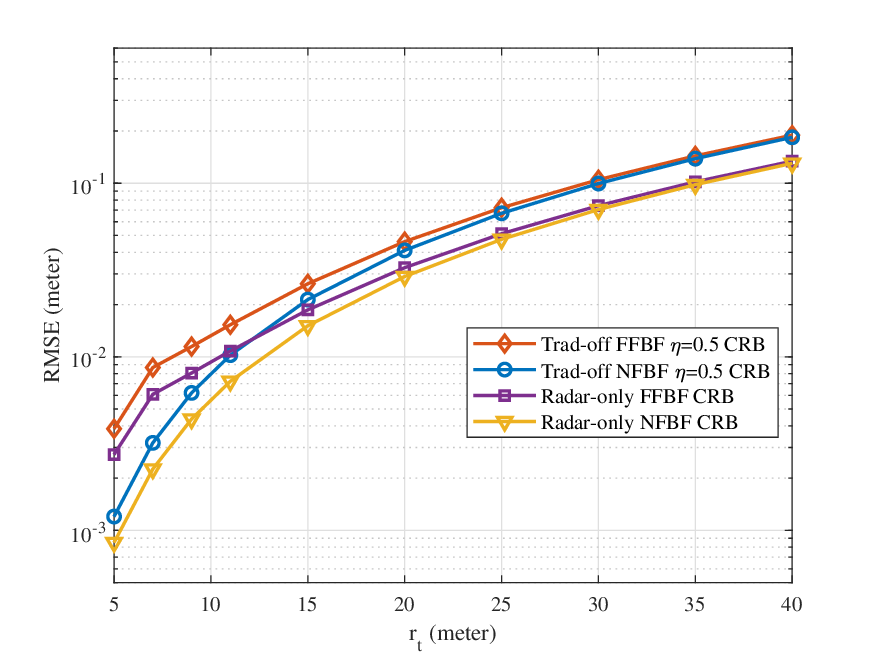}}
   \subfigure[Angle estimation.]{{\label{crb11}}\includegraphics[width=1\linewidth]{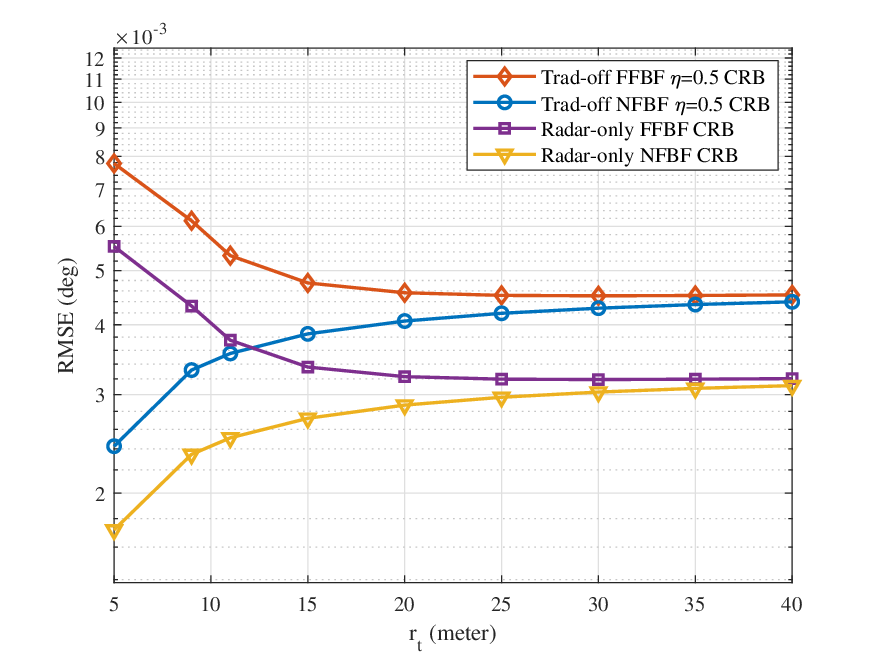}} 
  \caption{Effect of target distance on parameter estimation performance.}
  \label{crb222}
\end{figure}

Fig. \ref{crb222} demonstrates the effect of target distance on beam sensing performance. As the distance between the transmitter and receiver gets closer, the ISAC beam sensing performance of the proposed NFBF design improves. In contrast, as the distance gets farther, the performance of both the NFBF and FFBF designs tends to be similar. This is due to the significant near-field effect of the XL-array at closer distances. This observation inspires the idea that in future ISAC systems, we can first use a traditional far-field simple beam design to roughly determine the distance and then use the proposed NFBF for precise positioning and beam alignment if the distance is closer. This approach can effectively balance the trade-off between performance and complexity in ISAC systems.

\begin{figure}[!h]
  \centering
  \subfigure[$\hat{G}=100$.]{{\label{p1}}\includegraphics[width=1\linewidth]{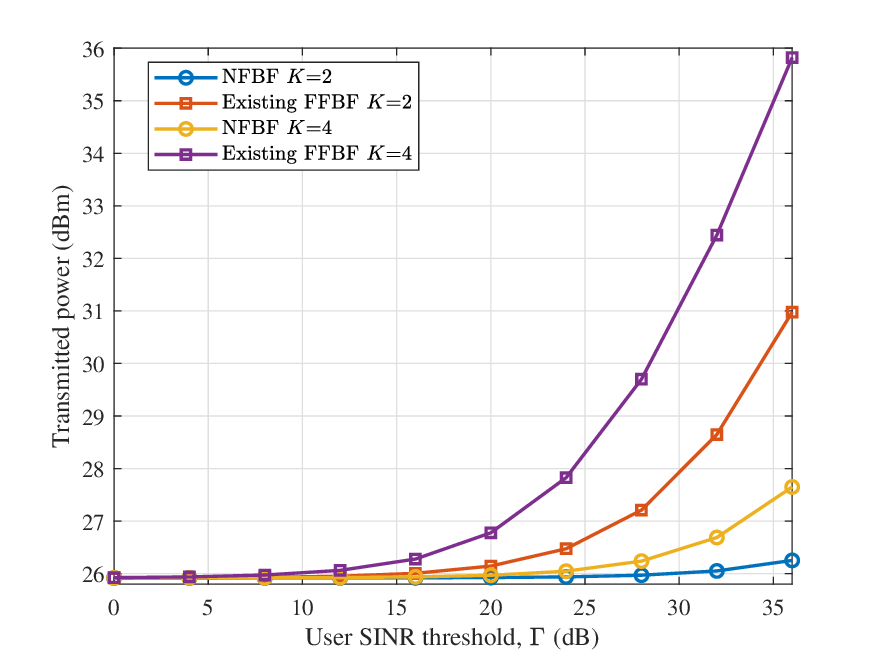}}
  \subfigure[$\Gamma=15$ dB.]{{\label{p2}}\includegraphics[width=1\linewidth]{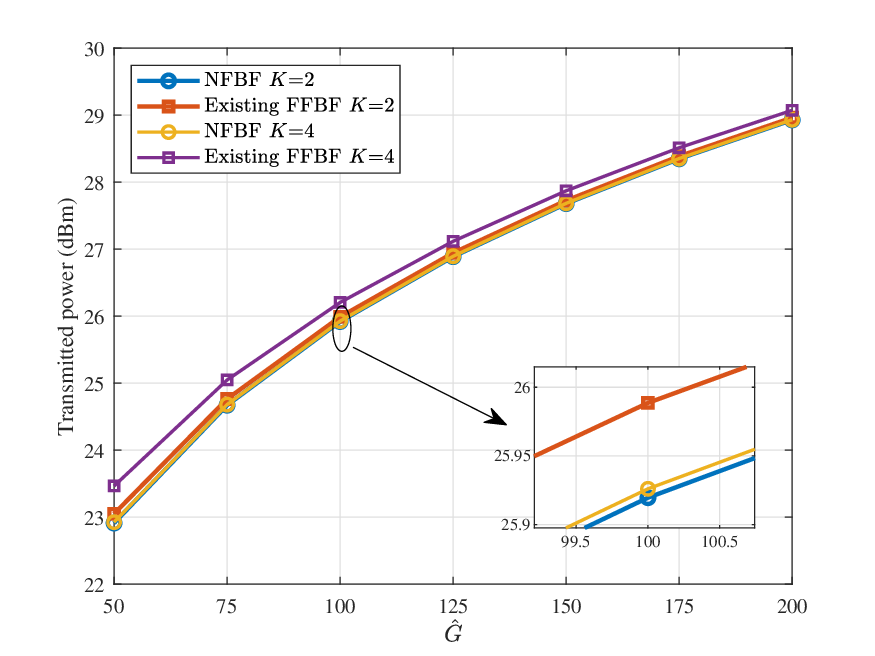}} 
    \caption{The transmitted power varies with the SINR threshold $\Gamma$ and target power $\hat{G}$. }
    \label{Power}
\end{figure}

\subsection{Power Consumption}
In the last simulation, we investigate the power performance of NFBF. We assume that each user has the same SINR requirement. In Fig. \ref{Power}, we compare the power consumption of ISAc beamforming occurring in the near field and far field as the user SINR threshold and target power $\hat{G}$ increases.
The power consumption becomes larger as the QoS requirements and the number of users increases. However, the power consumption of the NFBF is significantly lower than the other approach. This observation aligns with the previous analysis. For Fig. \ref{p1}, FFBF relies on scattering paths with higher pathloss, necessitating higher power to meet the SINR requirements. For Fig. \ref{p2}, FFBF has the gain loss shown in Lemma 1, and higher power is needed to compensate for that loss. This indicates that NFBF not only improves the beam sensing and communication performance of ISAC systems but also reduces the overall power consumption, which is a crucial factor for energy-efficient communication systems.

\section{Conclusions}
In this paper, we investigated the design and performance analysis of NFBF for ISAC systems. We began by establishing the model of electromagnetic (EM) near-field spherical wave and far-field plane wave and compared their differences to analyze the near-field beam focusing capability and far-field beam steering capability. We then derived the mathematical expression for the gain loss caused by the far-field steering vector mismatch in the near-field case.
Next, we encapsulated the properties of the near field into (\ref{eq4}) and formulated a weighted minimization problem for beam design, enabling a flexible trade-off between radar and communication objectives. However, due to the nonconvex nature of the optimization problems, we designed two algorithms to solve the problem. Simulation results showed that NFBF outperformed conventional FFBF, indicating the importance of studying near-field beamforming in extremely large-scale ISAC systems in-depth.
Inspired by this, we compared the power consumption of NFBF for the same QoS, and the results showed that NFBF effectively reduced the transmitting power. Additionally, the simulation results illustrated the importance of modeling the near field of XL-array correctly.

\begin{appendices}
\section{Proof of Lemma 1}\label{A1}
\begin{IEEEproof}
According to (\ref{eq4}), it is easy to get that $\mathbf{a}^H(r,\theta)\mathbf{a}(r,\theta)=N$,
Then we have the $n$-th element of $\mathbf{a}(\theta)$
\begin{equation}
{a}_n(\theta)=e^{j\pi\delta_n\sin{\theta}},
\end{equation} and the $n$-th element of $\mathbf{a}(r,\theta)$
\begin{equation}\label{an}
a_n(r,\theta)=e^{-j\frac{2\pi(r_n-r)}{\lambda}}.
\end{equation}
The phase term in equation (\ref{an}) is approximated by a quadratic function obtained from a second-order Taylor expansion of the range equation,
\begin{equation}
\begin{split}
    r_n&=r\sqrt{1+\frac{(\delta_nd)^2}{r^2}-\frac{2\delta_nd\sin{\theta}}{r}}\\
    &\approx r\left(1+\frac{\delta_n^2d^2\cos^2{\theta}}{2r^2}-\frac{\delta_nd\sin{\theta}}{r}\right)\\
    &=r+\frac{\delta_n^2d^2\cos^2{\theta}}{2r}-\delta_nd\sin{\theta},
\end{split}
\end{equation}
which is called the Fresnel approximation \cite{8736783} and has been proved in \cite{Selvan2017FraunhoferAF} that the second-order expansion is accurate enough in the near-field region. 
Then (\ref{an}) can be written as
\begin{equation}
    a_n(r,\theta)=e^{-j\pi\left(\frac{\delta_n^2d^2\cos^2{\theta}}{r\lambda}-\delta_n\sin{\theta}\right)}.
\end{equation}
Then we have
\begin{equation}
\begin{split}\label{747}
    |\mathbf{a}^H(\theta)\mathbf{a}(r,\theta)|&=\left|\sum_{n=0}^{N-1}{a}_n^H(\theta){a}_n(r,\theta)\right|\\
    &=\left|\sum_{n=0}^{N-1}e^{-j\pi\frac{\delta_n^2\lambda\cos^2{\theta}}{4r}}\right|.
\end{split}
\end{equation}

Assume that $N$ is even, then $\delta_n^2=\delta_{N-1-n}^2$ and $|\mathbf{a}^H(\theta)\mathbf{a}(r,\theta)|= \left|\sum_{n=0}^{\frac{N}{2}-1}2e^{-j\pi\frac{\delta_n^2\lambda\cos^2{\theta}}{4r}}\right|$ can be obtained. Denote that $g_n=e^{-j\pi\frac{\delta_n^2\lambda\cos^2{\theta}}{4r}}, n=0,1\cdots,\frac{N}{2}-1$, we can rewrite (\ref{747}) as
\begin{equation}\label{100}
\begin{split}
    |\mathbf{a}^H(\theta)\mathbf{a}(r,\theta)|&=\sqrt{\left(\sum_{n=0}^{\frac{N}{2}-1}2g_n\right)\left(\sum_{n=0}^{\frac{N}{2}-1}2g_n\right)^*}\\
    &=\sqrt{4\sum_{i=0}^{\frac{N}{2}-1}\sum_{j=0}^{\frac{N}{2}-1}g_ig_j^*}.
\end{split}
\end{equation}
Moreover, due to the conjugate relationship, we can calculate that $g_ig_i^*=1$ and 
\begin{equation}\label{110}
\begin{split} 
    g_ig_j^*+g_jg_i^*&=e^{j\pi\frac{(\delta_j^2-\delta_i^2)\lambda\cos^2{\theta}}{4r}}+e^{j\pi\frac{(\delta_i^2-\delta_j^2)\lambda\cos^2{\theta}}{4r}}\\
    &=2\cos{\frac{(\delta_i^2-\delta_j^2)\lambda\pi\cos^2{\theta}}{4r}},~~\forall i<j.
\end{split}
\end{equation}
Substituting (\ref{110}) into (\ref{100}) yields
\begin{equation}
    |\mathbf{a}^H(\theta)\mathbf{a}(r,\theta)|=\sqrt{2N+\sum_{i=0}^{\frac{N}{2}-2}\sum_{j=1,j>i}^{\frac{N}{2}-1}8\cos{\frac{(\delta_i^2-\delta_j^2)\lambda\pi\cos^2{\theta}}{4r}}}.
\end{equation}
 To further simplify the expression we encapsulate the coefficients $(\delta_i^2-\delta_j^2)$ in the upper triangular matrix $\mathbf{W}$,  mapping to the ordinal numbers $i, j$ and numbering of matrix $\mathbf{W}$  elements $\{i,j\}\to \{i+1,j\}$, then $\mathbf{W}$ can be derived as 
 \begin{equation*}
    \mathbf{W}= \begin{bmatrix}
N-2 & 2N-6 & \dots  & \frac{N^2-2N}{4} \\
  0 &  N-4  &    & \vdots \\
  \vdots & & \ddots  & 6\\
  0 &  0 &   \dots     & 2
\end{bmatrix}.
 \end{equation*}
The final form is expressed as
\begin{equation}\label{102}
   |\mathbf{a}^H(\theta)\mathbf{a}(r,\theta)|= \sqrt{2N+8\sum_{i=1}^{\frac{N}{2}-1}\sum_{j=1}^{\frac{N}{2}-1}\cos\left(\frac{\lambda \pi}{4r}\cos^2{(\theta)}\mathbf{W}\right)_{i,j}}.
\end{equation}
Substituting (\ref{102}) into (\ref{gainloss}), (\ref{obb}) is obtained, which completes the proof. 
\end{IEEEproof}

\section{Derivation of FIM MATRIX}\label{A2}
Firstly, we have
\begin{align}
    \frac{\partial \mathbf{u}}{\partial \boldsymbol{\phi}} &=\left[\beta \operatorname{vec}\left(\dot{\mathbf{G}_{\phi_1}}\mathbf{X}\right), \beta \operatorname{vec}\left(\dot{\mathbf{G}_{\phi_2}}\mathbf{X}\right)\right],\\
    \frac{\partial \mathbf{u}}{\partial \boldsymbol{\beta}}&=\operatorname{vec}\left(\mathbf{GX}\right)\left[1, j\right].
\end{align}

Then, $\mathbf{J}_{\boldsymbol{\phi} {\boldsymbol{\phi}}}$
can be further partitioned as
\begin{equation}
    \mathbf{J}_{\boldsymbol{\phi} {\boldsymbol{\phi}}}=\left[\begin{array}{ll}
    \mathbf{J}_{{\phi_1} {\phi_1}} & \mathbf{J}_{{\phi_1} {{\phi_2}}} \\
    \mathbf{J}_{{\phi_1} {{\phi_2}}}^{T} & \mathbf{J}_{{{\phi_2}} {\phi_2}}
    \end{array}\right].
\end{equation}

According to (\ref{eq35}), the entries  $\mathbf{J}_{{\phi_l} {\phi_p}}, \forall l, p \in\{1, 2\}$ can be
calculated as follows:

\begin{align}
     J_{{\phi}_l {\phi}_p}&=\frac{2}{\sigma_{w}^{2}} \operatorname{Re}\left\{\beta^{*} \operatorname{vec}\left(\dot{\mathbf{G}}_{\phi_{l}} \mathbf{X}\right)^{H} \beta \operatorname{vec}\left(\dot{\mathbf{G}}_{\phi_{p}}\mathbf{X}\right)\right\}\nonumber\\
    &=\frac{2|\beta|^2}{\sigma_{w}^{2}} \operatorname{Re}\left\{ \operatorname{tr}\left( \dot{\mathbf{G}}_{\phi_{p}}\mathbf{X}\mathbf{X}^H \dot{\mathbf{G}}_{\phi_{l}}^H\right)\right\}\nonumber\\
    &=\frac{2|\beta|^2L}{\sigma_{w}^{2}} \operatorname{Re}\left\{ \operatorname{tr}\left(  \dot{\mathbf{G}}_{\phi_{p}}\mathbf{R}_x\dot{\mathbf{G}}_{\phi_{l}}^H\right)\right\}.
\end{align}
Next, the matrices $ \mathbf{J}_{\boldsymbol{\phi} \tilde{\boldsymbol{\beta}}}$ and $ \mathbf{J}_{\tilde{\boldsymbol{\beta}} \tilde{\boldsymbol{\beta}}}$ are derived as follows
\begin{align}
    \mathbf{J}_{\boldsymbol{\phi} \tilde{\boldsymbol{\beta}}}&=\frac{2}{\sigma_{w}^{2}} \operatorname{Re}\left\{\left[\begin{array}{ll} 
    \beta^{*} \operatorname{vec}\left(\dot{\mathbf{G}_{\phi_1}}\mathbf{X}\right)^H\\
    \beta^{*} \operatorname{vec}\left(\dot{\mathbf{G}_{\phi_2}}\mathbf{X}\right)^H\end{array}\right],
    \operatorname{vec}\left(\mathbf{GX}\right)\left[1, j\right]
    \right\}\nonumber\\
    &=\frac{2}{\sigma_{w}^{2}} \operatorname{Re}\left\{\left[\begin{array}{ll} 
    \beta^{*} \operatorname{tr}\left(\mathbf{GX}\mathbf{X}^H\dot{\mathbf{G}}_{\phi_1}^H\right)\\
    \beta^{*} \operatorname{tr}\left(\mathbf{GX}\mathbf{X}^H\dot{\mathbf{G}}_{\phi_2}^H\right)\end{array}\right]
    \left[1, j\right]
    \right\}\nonumber\\
    &=\frac{2L}{\sigma_{w}^{2}} \operatorname{Re}\left\{\left[\begin{array}{ll} 
    \beta^{*} \operatorname{tr}\left(\mathbf{G}\mathbf{R}_x\dot{\mathbf{G}}_{\phi_1}^H\right)\\
    \beta^{*} \operatorname{tr}\left(\mathbf{G}\mathbf{R}_x\dot{\mathbf{G}}_{\phi_2}^H\right)\end{array}\right]
    \left[1, j\right]
    \right\},\label{eq48}\\
    \nonumber \\
    \mathbf{J}_{\tilde{\boldsymbol{\beta}} \tilde{\boldsymbol{\beta}}}
    &=\frac{2}{\sigma_{w}^{2}} \operatorname{Re}\left\{\left(\operatorname{vec}\left(\mathbf{GX}\right)\left[1, j\right]\right)^H\left(\operatorname{vec}\left(\mathbf{GX}\right)\left[1, j\right]\right)\right\}\nonumber\\
    &=\frac{2}{\sigma_{w}^{2}} \operatorname{Re}\left\{\left[1, j\right]^H\left[1, j\right]\operatorname{tr}\left(\mathbf{GX}\mathbf{X}^H\mathbf{G}^H\right)\right\}\nonumber\\
    &=\frac{2L}{\sigma_{w}^{2}}\mathbf{I}_2 \operatorname{tr}\left(\mathbf{G}\mathbf{R}_x\mathbf{G}^H\right).\label{eq49}
\end{align}
\section{Derivation of (41)-(43)}\label{A3}
First, according to (\ref{eq48}) (\ref{eq49}), we have
\begin{equation}
    \mathbf{J}_{\tilde{\boldsymbol{\beta}} \tilde{\boldsymbol{\beta}}}^{-1}=\frac{\sigma_{w}^{2}}{2L\operatorname{tr}\left(\mathbf{G}\mathbf{R}_x\mathbf{G}^H\right)}\mathbf{I}_2,
\end{equation}
and

\begin{equation}
\begin{split}
\label{eqJ}
    &\mathbf{J}_{\boldsymbol{\phi} \tilde{\boldsymbol{\beta}}}\mathbf{J}_{\boldsymbol{\phi} \tilde{\boldsymbol{\beta}}}^{T}=\left(\frac{2L}{\sigma_{w}^{2}}\right)^2\cdot\\&\left[\setlength{\arraycolsep}{1.5pt}\begin{array}{ll}
    \operatorname{Re}\left\{\beta^{*} \operatorname{tr}\left(\mathbf{G}\mathbf{R}_x\dot{\mathbf{G}}_{\phi_1}^H\right)\right\} &\ -\operatorname{Im}\left\{ \beta^{*} \operatorname{tr}\left(\mathbf{G}\mathbf{R}_x\dot{\mathbf{G}}_{\phi_1}^H\right)\right\}\\
    \operatorname{Re}\left\{\beta^{*} \operatorname{tr}\left(\mathbf{G}\mathbf{R}_x\dot{\mathbf{G}}_{\phi_2}^H\right)\right\}  & \ -\operatorname{Im}\left\{\beta^{*} \operatorname{tr}\left(\mathbf{G}\mathbf{R}_x\dot{\mathbf{G}}_{\phi_2}^H\right)\right\} \end{array}\right]\bf{\cdot}\\
    &\left[\setlength{\arraycolsep}{1.5pt}\begin{array}{ll}
    \operatorname{Re}\left\{\beta^{*} \operatorname{tr}\left(\mathbf{G}\mathbf{R}_x\dot{\mathbf{G}}_{\phi_1}^H\right)\right\} &\ \operatorname{Re}\left\{ \beta^{*} \operatorname{tr}\left(\mathbf{G}\mathbf{R}_x\dot{\mathbf{G}}_{\phi_2}^H\right)\right\}\\
    -\operatorname{Im}\left\{\beta^{*} \operatorname{tr}\left(\mathbf{G}\mathbf{R}_x\dot{\mathbf{G}}_{\phi_1}^H\right)\right\}  &\  -\operatorname{Im}\left\{\beta^{*} \operatorname{tr}\left(\mathbf{G}\mathbf{R}_x\dot{\mathbf{G}}_{\phi_2}^H\right)\right\} \end{array}\right].\\    
\end{split}
\end{equation}

For any two complex numbers $x=a+bi$ and $y=c+di$, we can obtain 
\begin{equation}
         \operatorname{Re}\{x\}\operatorname{Re}\{y\}+\operatorname{Im}\{x\}\operatorname{Im}\{y\}=ac+bd=\operatorname{Re}\left\{xy^{*}\right\}.
\end{equation}
Then (\ref{eqJ}) can be  simplified to (\ref{eq511})

\begin{figure*}[!t]
\normalsize
\setcounter{MYtempeqncnt}{\value{equation}}

\begin{equation}
\begin{split}
\label{eq511}
\mathbf{J}_{\boldsymbol{\phi} \tilde{\boldsymbol{\beta}}}\mathbf{J}_{\boldsymbol{\phi} \tilde{\boldsymbol{\beta}}}^{T}
&=\left(\frac{2L|\beta|}{\sigma_{w}^{2}}\right)^2\left[\setlength{\arraycolsep}{1.2pt}\begin{array}{ll}
    \left|\operatorname{tr}\left(\mathbf{G}\mathbf{R}_x\dot{\mathbf{G}}_{\phi_1}^H\right)\right|^2  &\quad  \operatorname{Re}\left\{ \operatorname{tr}\left(\mathbf{G}\mathbf{R}_x\dot{\mathbf{G}}_{\phi_1}^H\right)\operatorname{tr}\left(\mathbf{G}\mathbf{R}_x\dot{\mathbf{G}}_{\phi_2}^H\right)^{*}\right\}   \\\\
  \operatorname{Re}\left\{\operatorname{tr}\left(\mathbf{G}\mathbf{R}_x\dot{\mathbf{G}}_{\phi_2}^H\right)\operatorname{tr}\left(\mathbf{G}\mathbf{R}_x\dot{\mathbf{G}}_{\phi_1}^H\right)^{*}\right\}  &\quad  \left|\operatorname{tr}\left(\mathbf{G}\mathbf{R}_x\dot{\mathbf{G}}_{\phi_2}^H\right)\right|^2
\end{array}\right].
\end{split}
\end{equation}
\setcounter{equation}{\value{MYtempeqncnt}}
\hrulefill
\vspace*{4pt}
\end{figure*}

Then,
\setcounter{equation}{69}
\begin{equation}
\label{eq54}
\begin{split}
    \mathbf{J}_{\boldsymbol{\phi} \tilde{\boldsymbol{\beta}}} \mathbf{J}_{\tilde{\boldsymbol{\beta}} \tilde{\boldsymbol{\beta}}}^{-1} \mathbf{J}_{\boldsymbol{\phi} \tilde{\boldsymbol{\beta}}}^{T}&=\frac{\sigma_{w}^{2}}{2L\operatorname{tr}\left(\mathbf{G}\mathbf{R}_x\mathbf{G}^H\right)}\mathbf{J}_{\boldsymbol{\phi} \tilde{\boldsymbol{\beta}}}\mathbf{I}_2 \mathbf{J}_{\boldsymbol{\phi} \tilde{\boldsymbol{\beta}}}^{T}\\
    &=\frac{\sigma_{w}^{2}}{2L\operatorname{tr}\left(\mathbf{G}\mathbf{R}_x\mathbf{G}^H\right)}\mathbf{J}_{\boldsymbol{\phi} \tilde{\boldsymbol{\beta}}}\mathbf{J}_{\boldsymbol{\phi} \tilde{\boldsymbol{\beta}}}^{T}.
\end{split}
\end{equation}

Substituting (\ref{eq511}) and (\ref{eq54}) into (\ref{eq40}), it is easy to derive equation (\ref{CRB}).

Specifically, when the distance to the target is known, i.e. when $r_t$ is a constant, we have $\dot{\mathbf{G}}_{\phi_1}=0$, so (\ref{CRB}) degenerates to a matrix where only the element in the lower right corner is non-zero, (\ref{eq42}) is easily obtained. Similarly, when the angle of the target is known, i.e. $\theta_t$ is a constant, we have $\dot{\mathbf{G}}_{\phi_2}=0$, so (\ref{CRB}) degenerates to a matrix where only the element in the top left corner is non-zero, (\ref{eq43}) is easily obtained.
\end{appendices}

\bibliographystyle{IEEEtran} 
\bibliography{IEEEabrv,bib}
\end{document}